\documentclass[aps,twocolumn,nobibnotes,altaffilletter,amsmath,amssymb,amsfonts]{revtex4-1}
\usepackage[latin1]{inputenc}
\usepackage{graphicx}
\usepackage{amsmath}
\usepackage{latexsym}
\usepackage{epsfig}
\usepackage{subfigure}

\begin{document}

\title{Unveiling the complex glassy dynamics of square shoulder systems: \\simulations and theory}

\author{Gayatri Das$^1$, Nicoletta Gnan$^1$, Francesco Sciortino$^{1,2}$ and Emanuela Zaccarelli$^{2,1}$}

\affiliation{ $^1$Dip. di Fisica, Sapienza Universit\`a di Roma, P.le A. Moro 2, I-00185, Roma, Italy }
\affiliation{ $^2$CNR-ISC, Institute of Complex Systems, UoS Sapienza, Dip. di Fisica, Sapienza Universit\`a di Roma , P.le A. Moro 2,  I-00185, Roma, Italy }

\date{\today}
\begin{abstract}
We performed extensive molecular dynamics (MD) simulations, supplemented by Mode Coupling Theory (MCT) calculations, for the Square Shoulder (SS) model, a purely repulsive potential where the hard-core is complemented by a finite shoulder. For the one-component version of this model, MCT predicted [Sperl {\it et al.} Phys. Rev. Lett. {\bf 104}, 145701 (2010)] the presence of diffusion anomalies both upon cooling and upon compression and the occurrence of glass-glass transitions. In the simulations, we focus on a non-crystallising binary mixture, which, at the investigated shoulder width, shows a non-monotonic behaviour of the diffusion upon cooling but not upon isothermal compression.  In addition, we find the presence of a disconnected glass-glass line in the phase diagram, ending in two higher-order singularities. These points generate a logarithmic dependence of the density correlators as well as a subdiffusive behaviour of the mean squared displacement, although with the interference of the nearby liquid-glass transition. We also perform novel MCT calculations using as input the partial structure factors obtained within MD, confirming the simulation results. The presence of two hard sphere glasses, differing only in their hard core length, is revealed, showing that the simple competition between the two is sufficient for creating a rather complex dynamical behaviour.
\end{abstract}

\maketitle

\section{Introduction}
In the last decades, a lot of effort has been devoted to understand dynamical arrest  in soft matter systems. The pioneering investigations of hard-sphere (HS) colloids by Pusey and van Megen
 \cite{Pus87a,vanM93} have shown that the HS glass transition occurs at a colloidal packing fraction $\phi \approx 0.58$. This transition has been interpreted by  the ideal Mode Coupling Theory (MCT) \cite{goetze} for the glass transition. Despite suffering of a shift of the actual glass transition value, MCT provides a good description of the experimental data. The mechanism of arrest is explained in terms of the so-called `cage effect' { \cite{goetze,Gotzesperl}}, where particles at high densities become trapped by their nearest neighbours for an increasingly long time. This mechanism manifests itself in the form of a two-step decay of the density auto-correlation functions approaching the liquid-glass transition.

Subsequent investigations have focused on HS colloids in which an additional short-range attraction was added with the intent of mimicking the effective interaction (i.e. depletion) arising between colloids in suspension with non-absorbing polymers \cite{Pha02a}. A simple square-well (SW) attraction can be used to imitate these systems. MCT predictions \cite{Fab99a,Ber99a,Daw00a} for the 
dynamics of the SW model at high densities  revealed an intriguing behaviour. Indeed, when the range of the
 well width $\Delta$ is reduced down to a few percent of the particle diameter, a reentrant glass line is observed in the temperature-concentration phase diagram. This results in two different kind of glasses: a first glass (named repulsive glass), which is found at high temperature $T$, is the HS glass driven by the packing of particles, while a
 second glass (named attractive glass) is observed at low $T$, when energetic effects are dominant and particles remain caged in their attractive wells. In between the two glasses, at intermediate temperatures, a reentrant liquid region occurs. Therefore at the same concentration it is possible to go from one glass to the other by lowering $T$ and passing through a pocket of liquid states arising from the competition between energetic and entropic effects occurring at intermediate $T$ \cite{Sciortino2002b}. At even higher densities, a glass-glass line is observed, which terminates at an endpoint named higher order singularity. Associated to these multiple glasses and reentrant melting, MCT predicts the occurrence of anomalous dynamics, which results in a logarithmic (rather than two-step) decay of the density auto-correlation functions approaching the endpoint singularity, as well as in a subdiffusive behaviour of the particles mean-squared-displacement (MSD). Most of these predictions have been confirmed by several simulations \cite{Pue02a,Zac02a,ZaccarelliPNAS} and experiments \cite{Pha02a,Eck02a}. In particular, in simulations, it turns out to be useful to draw iso-diffusivity lines \cite{Fof02a} in the phase diagram. These lines maintain the same shape of the MCT glass line at all (sufficiently small) values of $D$, so that they provide a useful reference to establish whether a reentrance (and eventually associated anomalous dynamics) is present.

The successful predictions of MCT for the SW system paved the way for applications of the theory in a wide variety of soft matter systems. In particular, MCT has been used to describe the arrested behaviour of several, purely repulsive systems. Among these are star polymers \cite{Lik98c}, i.e. long polymer chains anchored onto a central core, where the number of chains (arms) varies the softness of the particles, bridging HS colloids (in the limit of very large arm number) to polymer chains (when the arm number is limited to 2). While one-component star polymer solutions only display a glass driven by packing of the stars \cite{Fof03a}, binary mixtures of stars of different arm numbers and sizes have been shown to display multiple glassy states through a combined effort of MCT, simulations and experiments \cite{MayerNature}. Despite indications that anomalous dynamics could be present in these systems \cite{MayerMacro}, a clear evidence from MCT predictions has not been provided. In contrast, a recent theoretical study of binary, size-asymmetric HS mixtures has reported the occurrence of higher-order singularities and a variety of different glasses \cite{Voigt2011}. 

Without necessarily turning to mixtures, it was recently realised that one-component systems with distinct length scales in the interaction potential (the so-called core softened models)  are also promising candidates for detecting thermodynamic and dynamic anomalies \cite{Jagla,Kumar05,core11,Ramp,repulsive,Gallo2012}. 
Among these, the simplest model is the square shoulder (SS) model, where the hard-core is complemented by an additional repulsive corona. This model has been used to describe the behavior of some metallic glasses \cite{cesium77} or complex materials like micellar \cite{Osterman07} or granular systems \cite{duran}, as well as primitive model of silica \cite{horbach08} and water \cite{Jagla}.

Recent MCT calculations reported the existence of multiple glass transitions also for the SS system both under compression and cooling \cite{Sperl2010}. A peculiar behaviour of the SS model, with no counterpart in other investigated systems, is the prediction of a disconnected glass-glass transition with two endpoint singularities for certain values of the shoulder width $\Delta$. Even though a rich phenomenology has been predicted for the SS system, numerical simulations aiming to confirm this behaviour have not been performed so far. In this work we provide an extensive and systematic characterisation of the SS model by means of event-driven molecular dynamics (MD) simulations in order to describe its dynamical behaviour. We examine the one-component system as well as a suitably chosen binary mixture which is considered in order to avoid crystallisation at high densities and low temperatures, and to probe a sufficiently slow dynamics.
The paper is organised as follows. In Section~\ref{sec:methods} we describe the simulation methods and provide a summary of MCT. Then in Section~\ref{sec:results} we report our main results in four different subsections:
in ~\ref{sec:diff} we discuss the behaviour of the self-diffusion coefficient calculated from the simulations and extract an ideal glass line using power-law fits of the data; in ~\ref{sec:MCT} we compare with existing MCT results and perform new calculations for the binary mixture currently under study to closely compare the theoretical results
with the simulations; in ~\ref{sec:hos} we then search for the existence of the predicted MCT higher order singularities;  in ~\ref{sec:fq} we report results for the non-ergodicity parameters obtained from theory and simulations to assess the types of the glasses that the system forms at various packing fractions and temperatures. Finally in Section~\ref{sec:discuss} we discuss our findings and provide some conclusions and perspectives.

\section{Methods: Simulations and Theory}
\label{sec:methods}
We study a $50:50$ mixture of $N=2000$ particles of species $A$ and $B$ interacting via pairwise SS potential
\begin{equation}
V_{ij}(r)=\begin{cases}\infty, & r<\sigma_{ij}\\
u_0, & \sigma_{ij}\leq r<(1+\Delta) \sigma_{ij}\\
0, & r\geq (1+\Delta)\sigma_{ij},\\ 
\end{cases}
\end{equation}
where $i,j = A,B$, $\sigma_{AA}$ and $\sigma_{BB}$ are the particles diameters (and $\sigma_{AB}=(\sigma_{AA}+\sigma_{BB})/2$), $\Delta\sigma_{ij}=0.15\sigma_{ij} $ are the shoulder widths, and $u_0=1$ is the shoulder height. The mass $m$ of both particles is chosen as unit mass, while $\sigma_{BB}$ and $u_0$ are the units of length and energy respectively. $T$ is measured in units of energy (i.e. $k_B$=1).

\begin{figure}[ht]
\begin{center}
\includegraphics[scale=0.3, angle=0]{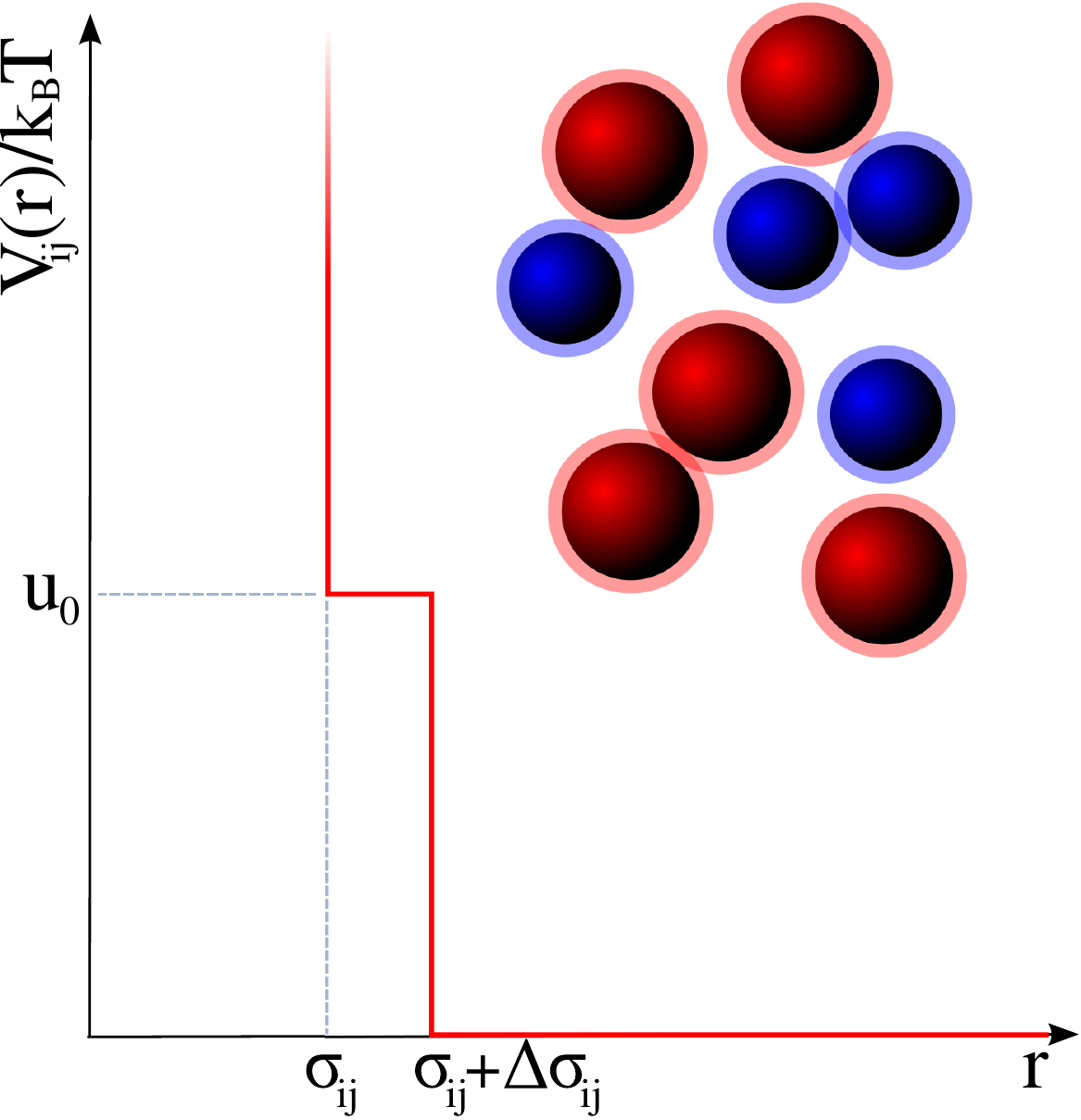}
\caption{Square shoulder potential for a generic additive binary mixture of species $i,j$. Here $\sigma_{ij}=(1/2)(\sigma_{i}+\sigma_{j})$ are the hard-cores, $\Delta\sigma_{ij}$ are the shoulder widths and $u_0=1$ is the shoulder height.}
\label{scheme}
\end{center}
\end{figure}

The size ratio between the two species is $\sigma_{AA}/\sigma_{BB}=1.2$.  We also study the simple monodisperse version of the SS system with the same width. However, the one-component system crystallises before it has actually entered a sufficiently slowed-down regime of dynamics, similarly to what generally observed for one-component glass-formers. 
The introduction of a small asymmetry in the size of the two species favors particles rearrangements at state points where the one-component easily crystallizes. This allows us to investigate states that are several orders of magnitude slower than those that are possible to explore in the monodisperse case. In this way, we  
 get as close as possible to the ideal glass line, which is defined as the locus of points in the packing fraction-temperature state diagram having diffusivity $D\rightarrow 0$. 

We perform  event-driven MD simulations of the system as a function of $T$ and packing fraction, defined as $\phi=(\pi/6)(\rho_A\sigma^3_A+\rho_B\sigma^3_B)$, being $\rho_i=N_i/L^3$, $L$ the edge of the cubic simulation box and $N_i$ the number of particles for each species. Simulations are performed in the canonical and microcanonical ensemble. For the desired packing fraction an initial configuration is generated randomly and the particles velocities are extracted from a Maxwell-Boltzmann distribution corresponding to the desired $T$. Then, the system is equilibrated by performing MD simulations in the canonical ensemble with appropriate rescaling of particle velocities. After the equilibration, for each state point investigated, NVE simulations are performed for times ranging from $t=10^{2}$ for the highly diffusive state points, to $t=10^{6}$ for the most viscous state points. For all the simulations $t$ is measured in units of $\sigma_{BB}(m/u_0)^{1/2}$.  

A first comparison with MCT results is possible by the calculation of iso-diffusivity lines, which typically preserve the shape of the ideal liquid-glass line. These lines, along which the self-diffusion coefficient $D$ is constant, are evaluated as follows. We first calculate the mean-squared displacement (MSD) $\langle r^2(t)\rangle$ of the particles at several $(\phi, T)$ and extract $D$ from its long-time limit behavior, using Einstein relation,
\begin{eqnarray}
D = \lim_{t\rightarrow \infty} \frac{\langle {r^2}(t) \rangle}{6t}.
\end{eqnarray}
Then we identify state points with the same $D$ and connect them by iso-$D$ lines. Repeating this procedure for lower and lower values of $D$, typically covering a few orders of magnitude in $D$, we can extrapolate the $D=0$-ideal glass line. To do this, we follow MCT predictions for $D$ that should go to zero at the ideal glass transition with a power-law dependence. This should apply independently on the chosen path, and hence both along an isotherm
\begin{equation}\label{eq:PowerD_phi}
D\sim\mid\phi-\phi_g(T)\mid^{\gamma(T)}
\end{equation}
\noindent and along an isochore
\begin{equation}\label{eq:PowerD_T}
 D\sim\mid T-T_g(\phi)\mid^{\gamma(\phi)}.
\end{equation}
\noindent Here $\phi_g$ and $T_g$ are, respectively, the critical values of the packing fraction and temperature at the ideal glass transition, while $\gamma$ is a non-universal exponent that is also determined by the theory.
By performing the power-law fits, following Eq.~(\ref{eq:PowerD_phi}) and (\ref{eq:PowerD_T}), we can then trace the locus of points in the ($\phi$, $T$) phase diagram for which $D\rightarrow0$. This line can be directly compared to the MCT glass line, as previously done for other systems \cite{Sci03a,Zac09a,MayerNature}. Indeed, MCT usually overestimates the tendency to form a glass, so that the two lines (numerical and theoretical) are always shifted by a certain amount in both $T$ and $\phi$. However, the shape of the two lines has been found, for all previously investigated systems, to be identical: this makes possible to establish an effective bilinear mapping between the two curves so that they scale on top of each other, as it has been done for the SW model \cite{sperl}. 
In the presence of singular state points, such as the  MCT higher order singularities \cite{Sperl0a}, the mapping procedure allows to estimate their exact location on the numerical phase diagram. Indeed, for the SW system, 
it was shown \cite{Sci03a} that one of such singularities does exist by performing ad-hoc simulations near this particular state point. In the present work, we aim to carry out a similar, detailed investigation for the SS system. 

To clarify some issues that will arise below,  we briefly summarize the main aspects of MCT here. 
MCT predicts the occurrence of a glass transition starting from a set of integro-differential equations for the density correlators $\Phi_q(t)=\langle\rho_{\bf{q}^*}(t)\rho_{\bf{q}}(0)\rangle/S(q)$ at different wave numbers $q$,
where $S(q)=\langle\rho_{\bf{q}^*}(0)\rho_{\bf{q}}(0)\rangle/N$ is the static structure factor and $\rho_{\bf q}(t)=\sum_{j=1}^{N_i} \exp[i{\bf q}\cdot {\bf r}_j(t)]$.
The MCT equations of motions (for Newtonian Dynamics)\footnote{The MCT equations can also be generalised to the case of Brownian Dynamics which is more realistic to describe colloidal suspensions. However, the long-time limit features and main predictions are not affected by the different microscopic dynamics.} read, in the one-component case, as
\begin{equation}\label{eq:mct}
\ddot{\Phi}_q(t)
+\Omega^2_q\Phi_q(t)+\Omega^2_q\int_0^t dt' m_q(t-t')\dot{\Phi}_q(t')=0
\end{equation}
\noindent where $\Omega^2_q=q^2k_BT/m S(q)$ is a characteristic frequency
and $m_q\equiv\mathcal{F}_q[\Phi_k(t)]$ is the memory kernel. 

Taking the long-time limit of Eq.~(\ref{eq:mct}) one obtains,
\begin{equation}\label{eq:fq}
f_q/(1-f_q)=\mathcal{F}_q[f_k],
\end{equation}
where $f_q=\lim_{t\rightarrow\infty}\Phi_q(t)$  is the so-called non-ergodicity parameter. 
$\mathcal{F}_q[f_k]$ is the Mode Coupling functional, which is bilinear in $f_q$
\begin{equation}\label{eq:MCT_Functional}
\mathcal{F}_q[f_k]=\frac{1}{2} \int \frac{d^3k}{(2\pi)^3}V_{{\bf q},{\bf k}} f_k f_{|{\bf q}-{\bf k}|},
\end{equation}  
where 
\begin{equation}\label{eq:Vertex}
V_{{\bf q},{\bf k}} \equiv S(q) S(k) S(|{\bf q}-{\bf k}|) \frac{\rho}{q^4}[{\bf q}\cdot{\bf k}c_k+{\bf q}\cdot({\bf q}-{\bf k})c_{|{\bf q}-{\bf k}|}]^2
\end{equation}
and $c_k=1/[1-\rho S(k)]$ is the direct correlation function.
At the glass transition MCT predicts that, for $t \rightarrow \infty$, the correlator does not decay to zero but reaches a finite plateau value. 

From Eq.~(\ref{eq:Vertex}) it is clear that the only inputs needed to solve Eq.~(\ref{eq:fq}) are the number density $\rho$ and the static structure factor $S(q)$ of the system. The latter can be obtained by solving the Ornstein-Zernike equation \cite{hansen} through the use of integral equations or it can be evaluated numerically from simulations. 

Besides the occurrence of a liquid-glass transition, under specific conditions, a system can display multiple glassy states, giving rise to the presence of glass-glass transitions in the kinetic phase diagram. These multiple glasses occur as bifurcations of the solutions of Eq.~(\ref{eq:fq}) upon variation of the control parameters. Across a glass-glass transition the non-ergodicity parameter jumps discontinuously between two non-zero values. This transition is found to terminate at an endpoint, named higher-order singularity, beyond which one can go from one glassy solution to the other continuously. The higher-order singularities can be of type $A_3$, when the two glasses coalesce already inside the glassy region, and of type $A_4$ when the two glasses merge also with the liquid solution right on top of the liquid glass line. The latter is a very special point  occurring at $(\phi^*, T^*, \Delta^*)$, which can be identified by finely tuning the value of the control parameter $\Delta$ \cite{Daw00a,Sperl2010}, and in its vicinity the form of the decay of $\Phi_q(t)$ is predicted to be unique.

Solving the full dynamical Eq.~(\ref{eq:mct}) close to any point on the liquid-glass transition, $\Phi_q(t)$ is found to follow a typical two-step decay. A first decay at short times corresponds to the characteristic time that particles employ to explore the cages formed by their nearest neighbors. A second decay occurs at longer time, and is characterized by the $\alpha$-relaxation time associated to the structural rearrangements necessary for restoring the ergodicity in the fluid. In between these two regimes, $\Phi_q(t)$ displays a characteristic plateau which is associated to the size of the cages in which particles are rattling before finally escaping.  It is well-accepted that the long-time relaxation of the correlators can be described by a stretched exponential
\begin{equation}\label{eq:stretch}
\Phi_q(t) \sim f_q \exp^{-(t/\tau_q)^{\beta_q}}
\end{equation}
where $f_q$, $\tau_q$ give an estimate respectively of the non-ergodicity parameter, the $\alpha$-relaxation time, while $\beta_q$ is the stretching exponent.

While no general analytic solution of the MCT equations is provided for  $\Phi_q(t)$, its asymptotic form close to 
the glass is known. On approaching the liquid-glass transition, the correlators are described by the Von Schweidler power-law decay \cite{goetze}. However, close to a higher order singularity, $\Phi_q(t)$ shows a peculiar logarithmic dependence:
\begin{equation}\label{eq:log}
\Phi_q(t)\sim f^{c}_q -h^{(1)}_q \ln(t/\tau_q) + h^{(2)}_q \ln^2(t/\tau_q).
\end{equation}
The parameters $f^{c}_q,h^{(1)}_q,h^{(2)}_q$ are the critical non-ergodicity parameter and critical amplitudes of first and second order in the expansion in $\ln(t)$ \cite{Gotzesperl}. It is found that a specific value of the wave vector $q^*$ exists, at which $h^{(2)}_q$ is zero, thus allowing for a pure logarithmic decay of the correlator to be observed. Hence, the correlators should display a characteristic concave (convex) shape for $q<q^*$ ($q>q^*$) in a logarithmic time scale.

For the SW system, higher-order singularities have been predicted and observed by numerical simulations and experiments. In particular, it was shown \cite{Daw00a,Daw02b} that MCT predictions in this case are robust upon the use of different closure relations such as mean spherical approximation (MSA) or Percus-Yevick (PY). Different closures only produce a shift of the glass transition lines with respect to each other.

For the SS system the situation appears to be more complex. Recent theoretical studies \cite{Sperl2010,Sperl2} have shown that for the same value of the shoulder width, the use of two different closures, namely
PY and Rogers-Young (RY), as input to the theory (from now on denoted as RY-MCT and PY-MCT respectively), provide qualitatively different results. While the liquid-glass line obtained within RY-MCT displays two reentrances (and hence diffusion minima and maxima) associated both to cooling (as in the SW system) and to compression, no reentrance is observed using PY-MCT. In the latter case, there is also no evidence of a glass-glass transition, while RY-MCT predicts two glass-glass lines each terminating in a higher order singularity. However, for the SS system at the investigated $\Delta$, RY is expected to be superior to PY in the description of its structural and thermodynamic properties. Therefore, one of the aims of this work will be also to assess the validity of PY-MCT or RY-MCT predictions in order to establish the correct scenario for the SS system while approaching the glass transition.

\section{Results}
\label{sec:results}
\subsection{Iso-diffusivity lines and ideal liquid-glass line from simulations} 
\label{sec:diff}
We start by reporting the behavior of $D$ along isothermal and isochoric cuts in the ($\phi$,$T$) phase diagram
in order to assess the presence of diffusion anomalies, perhaps like the ones observed in other 
core-softened potentials \cite{Kumar05,core11,Malescio0a}, conceptually similar to the SS system.
We remark that all shown data points do not crystallise and have reached a diffusive behaviour at long times, a condition necessary in order to extract $D$. In the following, we report results only for $A$ particles, because the behaviour of the $B$ particles is qualitatively the same due to the quasi-one-component nature of the mixture.

\begin{figure}
\includegraphics[width=0.4\textwidth]{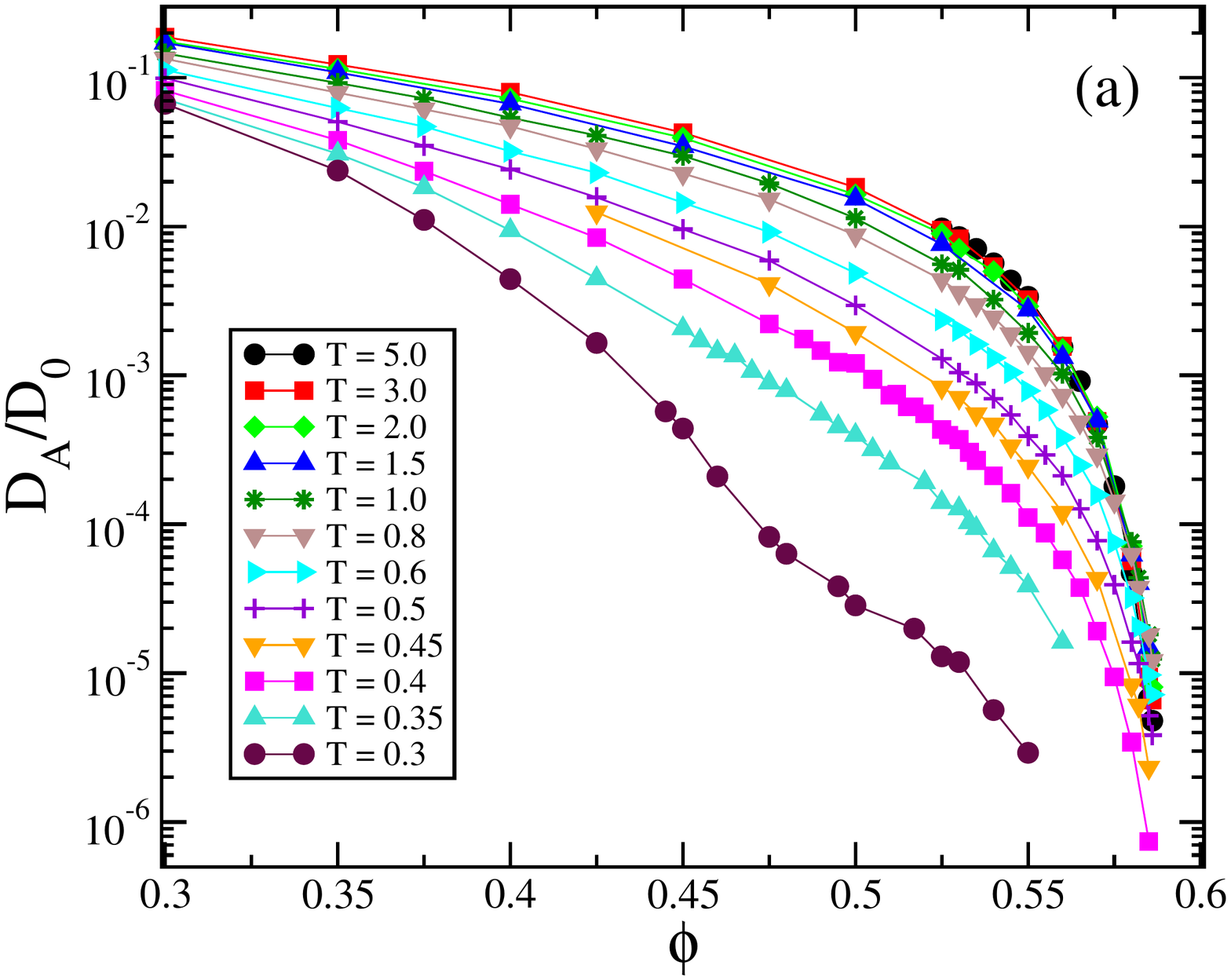}
\includegraphics[width=0.4\textwidth]{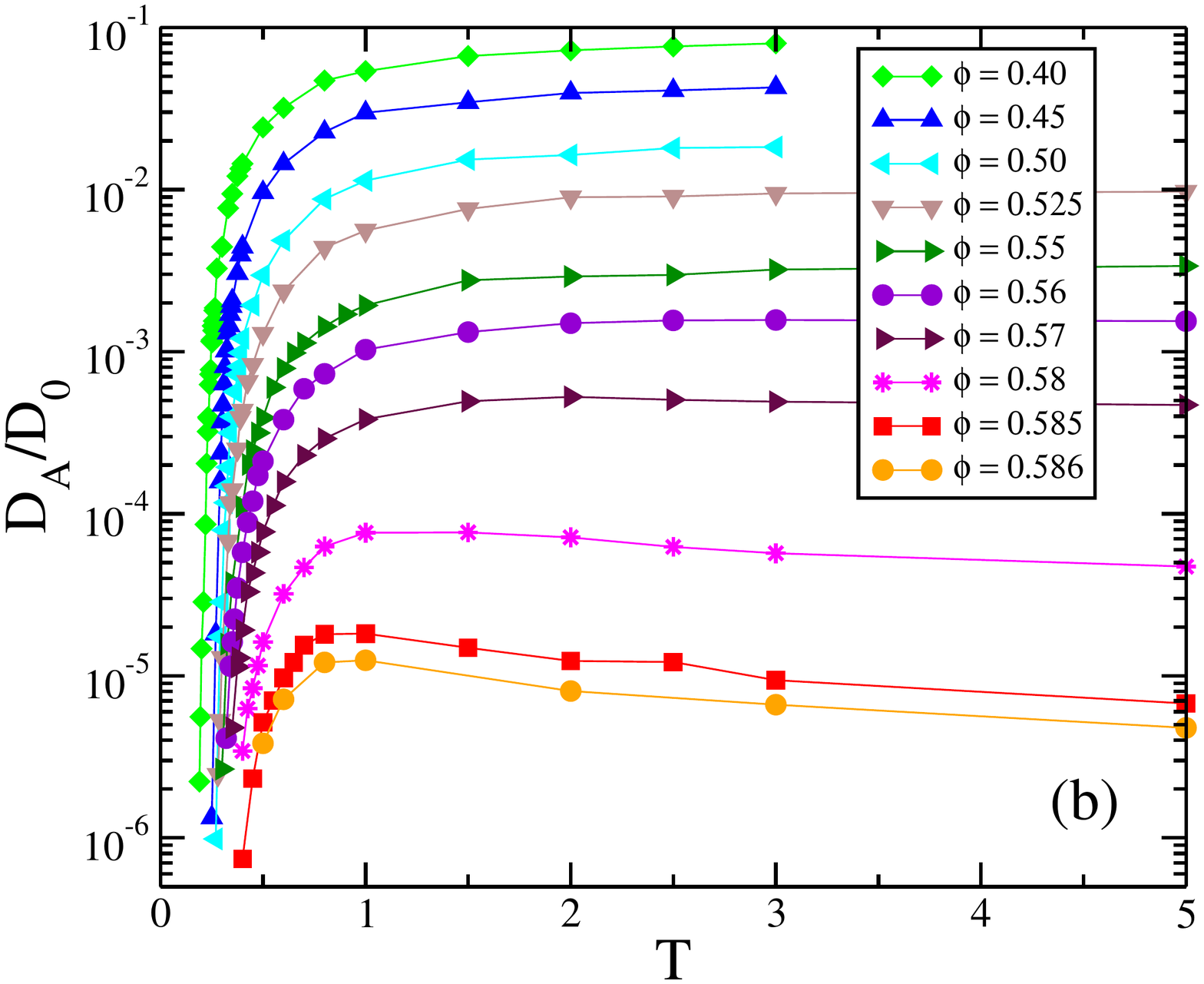}
\caption{Normalized diffusion coefficient $D_A/D_0$ as a function of (a) $\phi$ for several isotherms, as reported in the labels. At all investigated $T$ data show a monotonic decrease with increasing $\phi$, 
which clearly indicates the absence of diffusion anomalies associated to compression/expansion. A crossing of the data at high $\phi$ however signals the presence of a diffusivity maximum associated to cooling;
(b) $T$ for several isochores, as reported in the labels. }
\label{diffusion anomaly}
\end{figure}

Fig.~{\ref{diffusion anomaly}}(a) shows the normalized self-diffusion coefficient of A particles $D_A/D_0$
 as function of $\phi$ for several isotherms, varying from $T=5.0$ (close to the HS regime) to $T=0.3$ (where the shoulder effect becomes prominent). The normalization factor $D_0 = \sigma_{BB}{\sqrt{T/m}}$ is introduced to account for the $T$ dependence of the particles average velocities. 
 The behavior of $D_A$ with $\phi$ is similar along all studied isotherms: from the dilute limit $D_A$ decreases monotonically at all $T$. For not too low $T$,  the decrease becomes faster with increasing $\phi$ and is compatible with a power-law decay, as discussed below.  Notable exceptions are data points for $T\lesssim 0.35$: in these cases, a robust power-law dependence is not observed with decreasing $T$.
Indeed, at $T=0.3$ the data show a much wider range of decay. 
For a deeper investigation we have performed simulations for $T\leq 0.3$ and many adjacent $\phi$ with mesh 0.05 in the range $0.40 \leq \phi\leq 0.53$. This has allowed us to carefully check that the behavior of $D_A$ is strictly monotonic within numerical error at the studied $\Delta$ value of the SS model.
However, despite the absence of a diffusivity maximum, we are tempted to speculate that at these low $T$ the observed slower decrease of $D_A$ seems to be an effect of the competition between the two length scales in the potential.  Indeed, at low enough $\phi$ the system behaves as being composed of effective HS particles of diameter $\sigma+\Delta$, while with increasing $\phi$ the bare hard-core at $\sigma$ becomes dominant, thereby providing an intermediate non-trivial $\phi$-dependence of $D_A$. Such scenario does not exclude the presence of a non-monotonic behavior for different values of $\Delta$, that will be investigated in future studies.

Next, we investigate the behaviour of $D_A/D_0$ with $T$ along several different isochores, ranging between 
$\phi=0.40$ and $\phi= 0.59$. This is reported in Fig.~{\ref{diffusion anomaly}}(b). At lower $\phi$ the system 
does not reach a glass transition even for very low temperatures. Indeed, in this limit, it can be considered as an effective hard sphere of diameter $\sigma+\Delta$. Since hard spheres are expected to undergo a glass transition at $\phi_g^{HS} \sim 0.58$, we expect for the low-$T$ limit that the system becomes glassy for $\phi_g^{HS}[\Delta/(\sigma+\Delta)]^3 \sim 0.381$, in agreement with our simulations.

For packing fractions $0.40\le \phi\le 0.56$ the diffusion coefficient decreases monotonically with $T$. For $\phi\ge 0.57$, $D_A/D_0$ becomes non-monotonic: from the high $T$ limit, it initially increases and then decreases, giving rise to the presence of a diffusivity (local) maximum at intermediate $T$. This is also visible from the crossing of the high-$T$ data (at large $\phi$) in Fig.~{\ref{diffusion anomaly}}(a). The anomalous behavior of $D_A$ upon cooling is similar to that observed for the SW system at high enough packing fractions when the width of the well is of few percent of the particle size { \cite{Zac02a}}. 

Compiling all data from Figs.~\ref{diffusion anomaly}(a),(b) we are able to trace isodiffusivity lines in the phase diagram to be compared with the MCT glass lines. The monotonic (non-monotonic) behaviour of $D_A$ is reflected in the absence (presence) of a reentrance in the iso-$D_A$ lines.

\begin{figure}[h]
\includegraphics[width=0.4\textwidth]{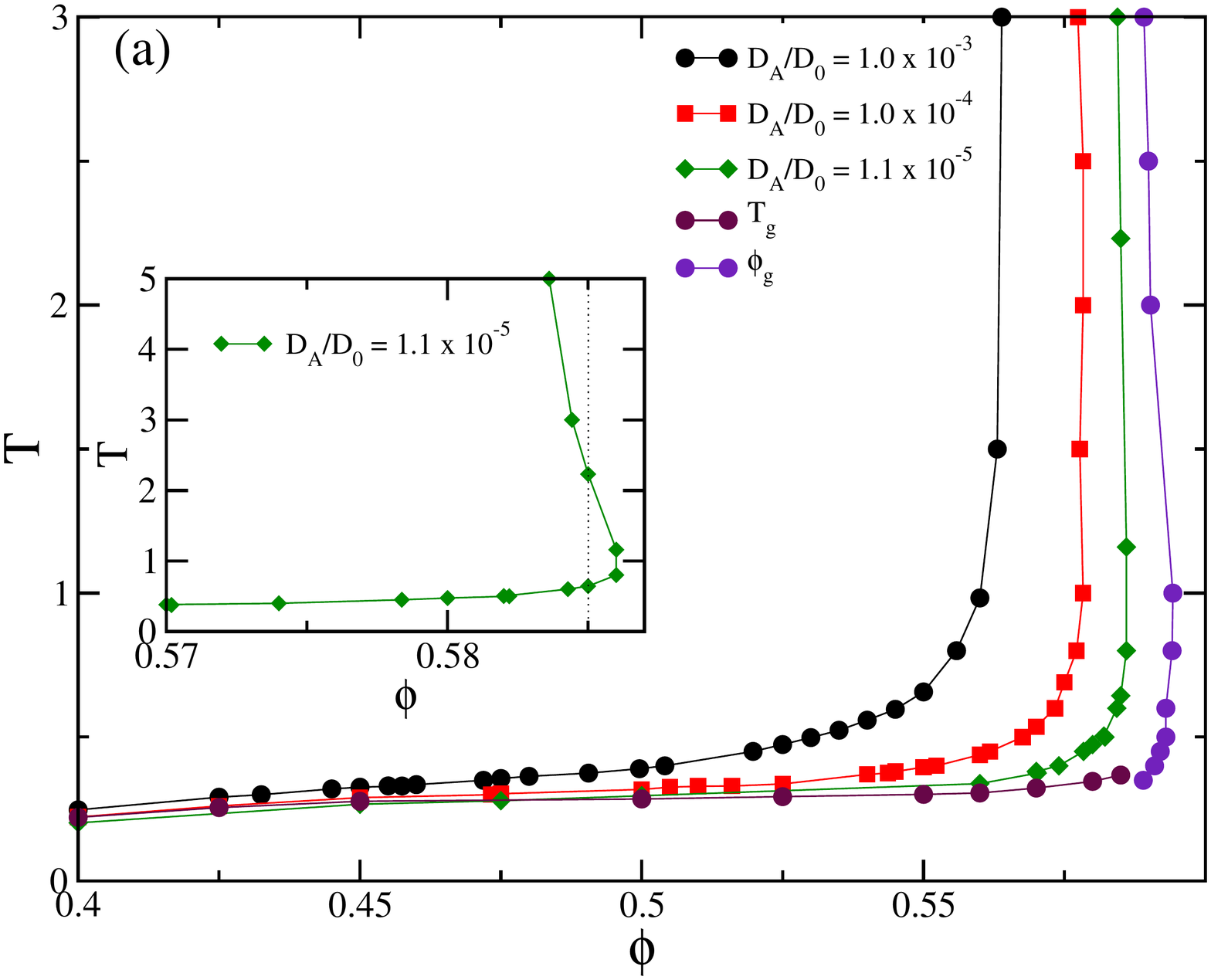}
\includegraphics[width=0.4\textwidth]{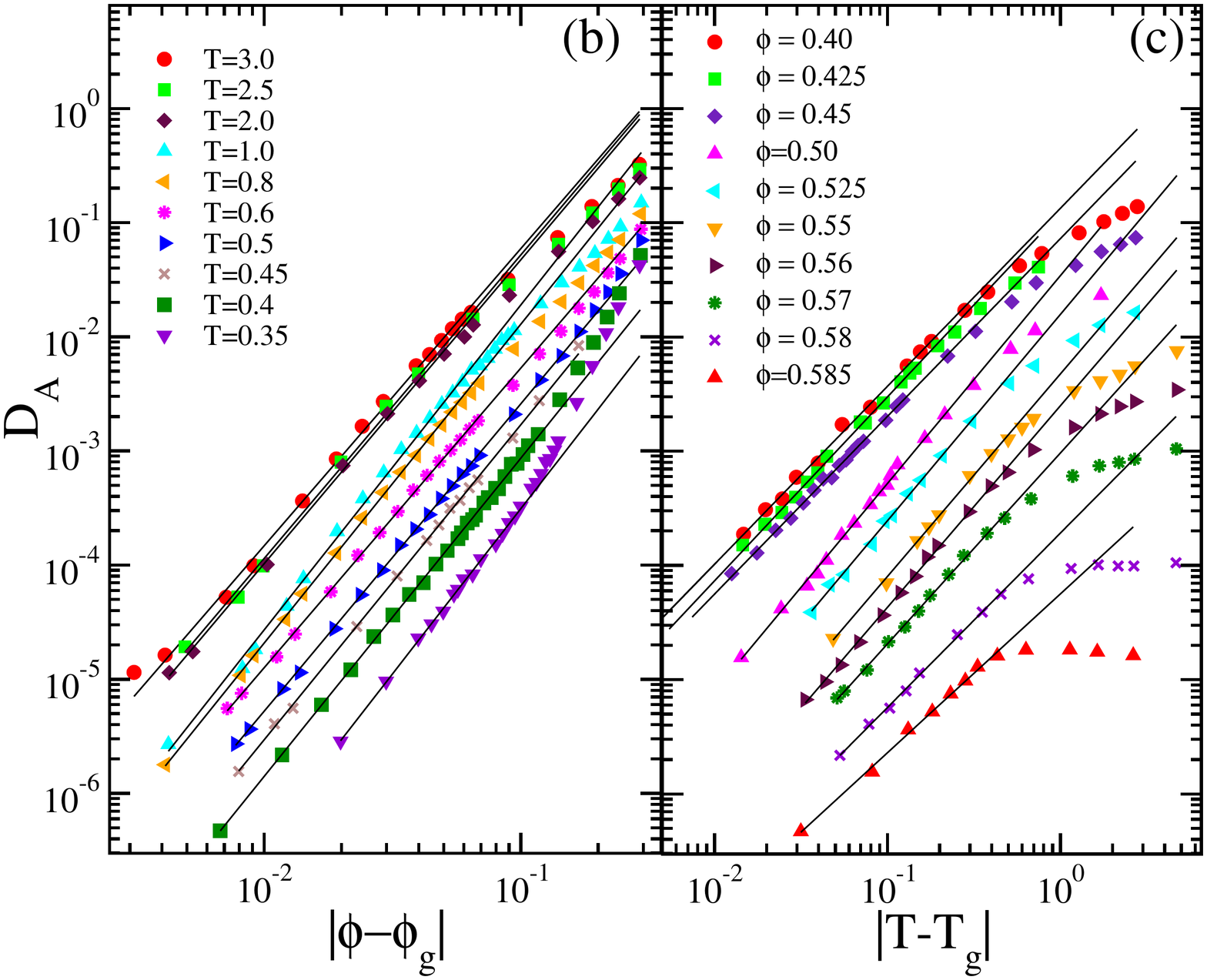}
\caption{(a) Isodiffusivity lines for $D_A/D_0=1.0\times10^{-3}, 1.0\times10^{-4}$ and $1.1\times10^{-5}$, as well as the extrapolated arrest ($D_A=0$) lines from the fits $D_A\sim |\phi-\phi_{g} (T)|^{\gamma(T)}$ along isotherms and $D_A\sim | T-T_{g} {\phi}|^{\gamma(\phi)}$ along isochores. The data display a reentrance in $T$ (inset), while no reentrance in $\phi$ is observed; (b) Power law fits along isotherms (left) and isochores (right). }
\label{isodiffusivity}
\end{figure}

Fig.~{\ref{isodiffusivity}}(a), shows the iso-diffusivity curves for three fixed values of normalised diffusion coefficients: $D_A/D_0=1.0\times10^{-3}, 1.0\times10^{-4}$ and $1.1\times10^{-5}$. Each curve is obtained by extrapolating from Fig.~{\ref{diffusion anomaly}}(a) and Fig.~{\ref{diffusion anomaly}}(b) a set of $i$ states $(\phi_i,T_i)$ having the same value of $D_A/D_0$. As discussed above, we do not observe a reentrant behaviour along $\phi$ even  if we consider very low values of $D_A/D_0$. On the other hand, for $D_A/D_0 \sim 10^{-5}$ a  reentrance is observed along $T$, as highlighted in the inset of Fig.~{\ref{isodiffusivity}}(a). This is in agreement with the presence of a diffusivity maximum at high enough $\phi$ (Fig.~{\ref{diffusion anomaly}}(b)).

From this analysis we conclude that the liquid-glass line shows only a reentrance along $T$.
While the glass at high $T$ corresponds to a HS glass, the low-$T$ glass could have a different nature and its properties will be elucidated in the following.

As said above, the iso-$D_A$ lines are precursors of the ideal liquid-glass line, where $D_A\rightarrow 0$. We can extrapolate this line where the simulated system should undergo dynamical arrest by performing power-law fits of the diffusivity both along isochores and along isotherms, following the MCT predictions in Eq.(\ref{eq:PowerD_phi}) and Eq.(\ref{eq:PowerD_T}). The fits are shown in Fig.~\ref{isodiffusivity}(b).  In this way, we can extract both the transition values $\phi_g$ and $T_g$ where the system arrests and the associated power-law exponents 
$\gamma$. These values are reported  in Table~\ref{table1}. Note that no fits are performed for $T<0.35$ due to the fact that the data do not show a clear power-law behaviour. Also for low values of $\phi$ some deviations from power-law dependence are observed at low $T$.
\begin{table}
\begin{tabular}{| l | l | l | l | l | l | l | l |}
\hline
~~~$T$~~~&~~~$\gamma(T)$~~~&~~~$\phi_g$~~~&~~&~~~$\phi$~~~&~~~$\gamma(\phi)$~~~&~~~$T_g$~~~\\ 
\hline
~~$3.0$ & ~~$2.6$ & ~$0.589$ & ~~& ~~$0.40$ & ~~$1.56$ & ~$0.221$\\
~~$2.5$ & ~~$2.66$ & ~$0.589$ & ~~& ~~$0.425$ & ~~$1.61$ & ~$0.255$\\
~~$2.0$ & ~~$2.66$ & ~$0.59$ & ~~&~~$0.45$ & ~~$1.56$ & ~$0.277$\\ 
~~$1.0$ & ~~$2.84$ & ~$0.594$ & ~~&~~$0.50$ & ~~$1.81$ & ~$0.285$\\
~~$0.8$ & ~~$2.79$ & ~$0.594$ & ~~&~~$0.525$ & ~~$1.81$ & ~$0.293$\\
~~$0.6$ & ~~$2.62$ & ~$0.593$ & ~~&~~$0.55$ & ~~$1.77$ & ~$0.301$\\
~~$0.5$ & ~~$2.69$ & ~$0.593$ & ~~&~~$0.56$ & ~~$1.76$ & ~$0.306$\\
~~$0.45$ & ~~$2.76$ & ~$0.592$ & ~~&~~$0.57$ & ~~$1.67$ & ~$0.323$\\
~~$0.4$ & ~~$2.78$ & ~$0.591$ & ~~&~~$0.58$ & ~~$1.52$ & ~$0.346$\\
~~$0.35$ & ~~$2.89$ & ~$0.589$ & ~~& ~~$0.585$ & ~~$1.39$ & ~$0.368$\\
\hline
\end{tabular}
\caption{Extrapolated values of $\gamma(T)$, $\phi_g$, $\gamma(\phi)$ and $T_g$ obtained from fitting data of Fig.~\ref{isodiffusivity} (a) and (b) with MCT predictions of Eqs.~(\ref{eq:PowerD_phi},\ref{eq:PowerD_T})  for the diffusion coefficient $D_A$. Error bars of the fit parameters typically amount to a few percent for the values of $\phi_g$  and $T_g$, while the $\gamma$ exponents can vary systematically over different fit intervals, so they should be taken with caution.}
\label{table1}
\end{table}

The resulting ideal glass line is also shown in Fig.~\ref{isodiffusivity}. As expected the two branches, extrapolated by the different paths, merge continuously in the high-$\phi$, low-$T$ region of the phase diagram and confirm the shape of the isodiffusivity lines. We note that the power-law exponents obtained along each isotherm are consistent with previous estimates for HS or SW systems. 
On the other hand, for the fits along isochores the $\gamma$ exponents are systematically lower, at times going below the lowest limit predicted by MCT \cite{goetze}.  However, the values of $\gamma$ obtained from the fits should be taken with caution due to the significant variation of results upon change of the chosen fit interval and relative distance to the transition. Nonetheless, the values of $\phi_g$ and $T_g$ extracted in the same way show only little changes (of the order of a few percent), and hence they are robust.

We note that power-law fits along isochores could not be performed in most systems with isotropic potentials, where the exploration of the low-$T$ region is preempted by intervening phase separation. For systems with directional interactions where phase separation is suppressed by using a limited valence \cite{EZ2006,sphericaljcp2}, at low $T$ bonding is the dominant mechanism of arrest so that the dynamics is dominated by an Arrhenius (strong) behaviour \cite{Boh93a}. Here, however, we do not find evidence of an Arrhenius dependence even at very low $T$ in the investigated window of densities, suggesting that the system remains power-law (fragile).  Indeed, in the SS system temperature does not induce bonding, but rather has an effect on the excluded volume of  particles by changing the effective diameter. In this sense, the arrest at low $T$ remains of the same kind of the HS glass, so that a similar behaviour (fragile) is then expected throughout the phase diagram. However, it is then legitimate to ask, given that the nature of the glass transition remains the same, whether a simple competition between two length scales is capable to generate higher-order singularities as 
those predicted by MCT and related glass-glass transitions.
 
\subsection{Comparison with old and new MCT results: 
role of the input structure factors and mapping to simulations}
\label{sec:MCT}
\begin{figure}[ht]
\includegraphics[scale=0.3, angle=0]{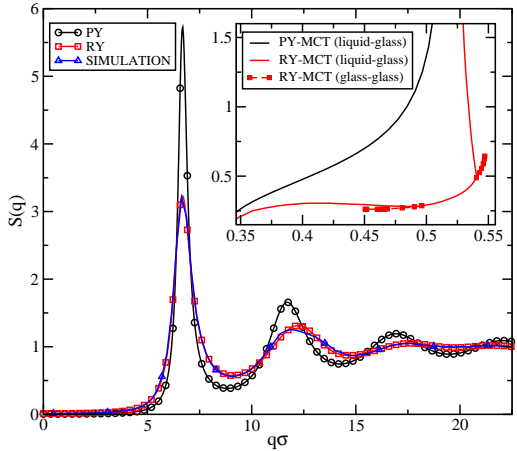}
\caption{Static structure factors for a monodisperse SS system at $T=0.5,\phi=0.45$ calculated by MD simulations as well as solving the Ornstein-Zernike equation within  Rogers-Young (RY)  and Percus-Yevick (PY) closures. Inset:
MCT results for the liquid-glass and glass-glass lines  using PY and RY. }
\label{fig:sq}
\end{figure}

We now compare the MD simulation results for the ideal glass line with MCT predictions.
While a mismatch of $\phi_g$ and $T_g$ values is expected for the theoretical and numerical glass lines, the two should share the same shape, as previously observed for a variety of glass-forming systems  \cite{Fof02a,Fof03a,Kumar05,Zac09a,MayerMacro}. However, when referring to the RY-MCT calculations for $\Delta=0.15$ \cite{Sperl2010}, it is immediate to notice that while the numerical curve shows only one reentrance in $T$, the RY-MCT results display two of them, as shown in the inset of Fig.~\ref{fig:sq}. No reentrance is conversely observed for PY-MCT\footnote{Courtesy of M. Sperl}.
Under this situation, we cannot perform a consistent mapping as previously done for SW systems \cite{Sci03a}, because the difference in the shape of the liquid-glass lines cannot be taken into account by a simple rescaling procedure. This matter thus deserves further investigation.

For understanding the difference between theory and MD results, at first we investigate the reliability of the different closures employed for producing the input $S(q)$ entering in MCT. In Fig.~\ref{fig:sq} $S(q)$ evaluated within RY and PY closures is shown together with that calculated directly from simulations of the monodisperse SS system for a representative state point. As expected, $PY$ provides a rather poor estimate of $S_q$, since the height of its peaks and its amplitudes do not agree with those of the $S_q$ evaluated from MD, while RY reproduces reasonably well the simulation results, as previously found for other repulsive potentials \cite{Lang99,Kumar05,Rheo}. 
The good agreement between RY and MD $S(q)$ is found for all studied state points. However, this comparison is limited to the region of the phase diagram where the monodisperse system does not crystallize.  
The quality of the input structure factors is reflected in the better agreement of RY-MCT with the simulation iso-diffusivity lines. It is therefore natural from now on, to refer to RY-MCT results as the relevant theoretical predictions for the system. 

However, although the shape of the RY-MCT liquid-glass line is more similar to the MD results, with at least one reentrance recovered, there is still a discrepancy between RY-MCT and MD simulations. In fact, both from the analysis of $D/D_0$ and from the iso-diffusivity lines we could not detect the presence of a second reentrance (i.e. a diffusion anomaly) along $\phi$ (at fixed low-$T$).

Given this situation, we performed additional MCT calculations explicitly incorporating the binary nature of the system under investigation.  To avoid to rely on a certain closure, we have used as inputs to the theory the partial structure factors evaluated from MD simulations $S_{ij}^{SIM}(q)$. Hence, we have solved the generalized version of long-time MCT equations (Eq.~\ref{eq:fq}) for a binary mixture \cite{gotze03} on a discretised grid of 1000 wave vectors up to a cut-off value of $q\sigma_{BB}=65$. This value is 
sufficient for the critical non-ergodicity parameters along the liquid-glass line to decay to zero.
In this way, we determine the liquid-glass and (if any) the glass-glass transition, and associated non-ergodicity parameters. 

\begin{figure}[ht]
\begin{center}
\includegraphics[scale=0.38, angle=0]{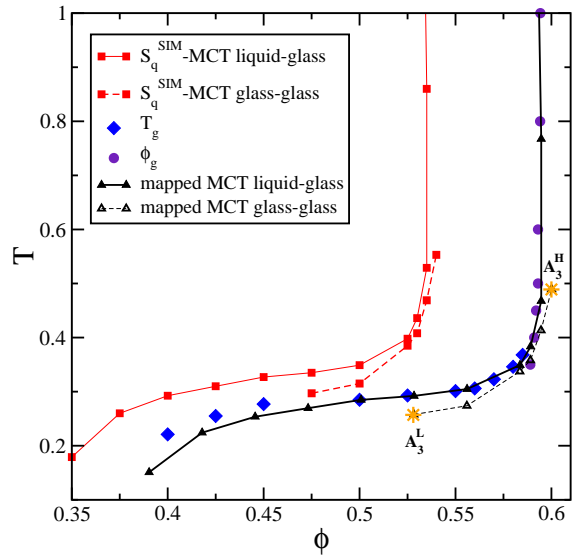}
\caption{MCT results for the binary mixture under study using the static structure factors calculated from simulations as input, labeled as $S_q^{SIM}-MCT$ liquid glass (filled squares) and glass-glass (open squares). Arrest curve drawn from $\phi_g$ (filled circles) and $T_g$ (filled diamonds) obtained from power-law fits of $D_A$ as in Fig.~\protect\ref{isodiffusivity}. Mapped MCT lines onto the arrest curve: liquid-glass (filled triangles) and glass-glass (open triangles). Stars are the two predicted higher order singularities $A_3^L$ and $A_3^H$. 
}
\label{simulation-mct}
\end{center}
\end{figure}

The resulting liquid-glass line is reported in Fig.~\ref{simulation-mct} together with the arrest line extrapolated from the fits of $D_A$. Despite the expected shift in the control parameters, it now appears that the new MCT results for the mixture are in full qualitative agreement with the simulation line, since the reentrance in $\phi$ is no longer present. 
We can now operate a bilinear transformation, as previously done for the SW system \cite{sperl,Sci03a},  to superimpose the MCT results onto the glass line obtained from simulations. The parameters are chosen via a best fit procedure, giving as a result
\begin{eqnarray}
\label{bilinear}
\phi &\rightarrow& 1.1046\phi+0.0038 \nonumber \\
T &\rightarrow& 0.9052T-0.0111
\end{eqnarray}
and the mapped glass lines are shown in Fig.~\ref{simulation-mct}. 

MCT calculations predict a `disconnected' glass-glass line, a scenario that was also present in the one-component RY-MCT, albeit for lower values of $\Delta$ \cite{Sperl2010}. Luckily, this glass-glass line lies just inside, but very close to the liquid-glass line so that signatures of the two $A_3$ endpoints are in principle detectable from simulations. Through our mapping, we can now estimate the location of the two singularities, that will be referred from now on as $A_3^L$ and $A_3^H$, indicating respectively the one at lower and higher $\phi$. We find $A_3^L=(\phi\sim0.53,T\sim0.26)$ and $A_3^H=(\phi\sim0.60,T\sim0.49)$. 

\subsection{Searching for higher-order singularities}
\label{sec:hos}
In this section we investigate the presence of the higher-order singularities predicted by MCT in the numerical phase diagram. To this aim we concentrate on distinct paths in the phase diagram that allow us to approach closely the two $A_3$ points. We recall however, that both points are buried within the glass region, hence they are not directly accessible in equilibrium; moreover the behaviour of the observables that we examine are influenced also by the presence of the nearby liquid-glass transition. In the following we will only concentrate on species $A$, but we stress that the qualitative behaviour is identical for type B particles.

We start by discussing the presence of $A_3^L$: we examine the dynamical behaviour of the system along the isochore $\phi=0.525$ with decreasing $T$. We recall that while the endpoint should be found at $T\sim 0.26$, the system becomes glassy according to MCT for $T\lesssim 0.28$ at this volume fraction. 

\begin{figure}[h]
\includegraphics[scale=0.3, angle=0]{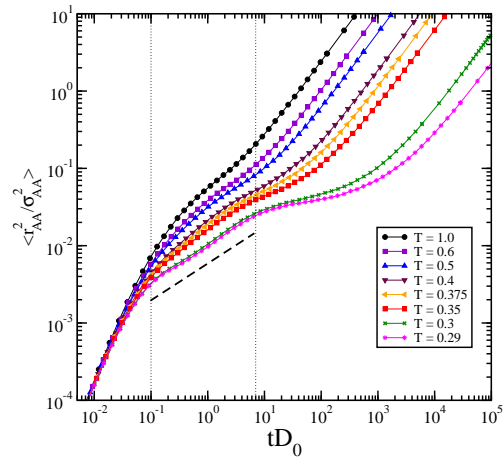}
\caption{MSD for a particles as a function of scaled time $tD_0$ for $\phi=0.525$ as a function of $T$, indicated in the labels.
The vertical dotted lines indicate as guides to the eye the regime of subdiffusive behaviour, which is highlighted by the dashed line ($\propto t^{0.5}$).}
\label{MSD-525}
\end{figure}

Fig.~\ref{MSD-525} shows the MSD for  A particles $\langle r^2_{AA}\rangle$ along this path. We observe that upon decreasing $T$ the system shows a peculiar slowing down.  Indeed, a characteristic subdiffusive behaviour at intermediate times ($0.1\lesssim tD_0 \lesssim 10$) is observed for $T < 0.4$. Hence, we observe a sort of three-step behaviour of the MSD: after the ballistic transient, subdiffusion takes place for roughly two decades in time, where $\langle r^2\rangle \sim t^{\alpha}$ with $\alpha \sim 0.5$.
 At long times the typical pattern of glass-forming systems takes place:  a plateau later followed by long-time diffusion. Indeed, when $T$ is very low, the system is approaching the liquid-glass transition, that manifests itself in the MSD as the emergence of the plateau. This is observed for $T \leq 0.3$ and occurs at $tD_0\sim 10^2$. The plateau height is found to be $\sim 0.04 \sigma^2_{AA}$. Its square root, which provides a measure of the cage or localisation length $l_0$ of the glass, turns out to be $\sim 0.2\sigma_{AA}$, roughly twice the typical HS cage length ($l_0^{HS}\sim 0.1 \sigma)$. Indeed, the packing fraction is significantly smaller than that of the HS ideal glass, due to the effect of the shoulder. While the long-time behaviour, and associated plateau, is controlled by the liquid-glass transition, the additional intermediate behaviour, which indicates the presence of a sub-diffusive regime, can be associated to the presence of the higher order singularity. 

\begin{figure}[h]
\includegraphics[scale=0.3, angle=0]{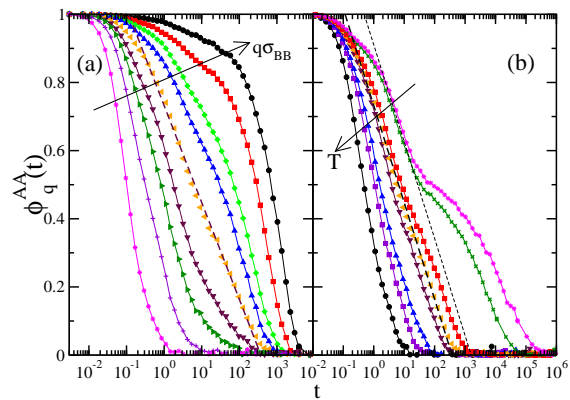}
\caption{The density autocorrelation functions $\Phi^{AA}_{q}(t)$ for $\phi=0.525$ as a function of time for (a) several wave vectors at $T=0.375$. From top to bottom, $q \sigma_{AA} = 1.88, 2.81, 4.68, 5.63, 7.5, 10.32, 13.12, 17.82, 28.13$. A concave-convex shape transition is observed around $q^*\sigma_{AA}\approx 7.0$ where the decay of $\Phi^{AA}_{q}(t)$ is almost purely logarithmic; (b) several $T$ at fixed wave vector $q=q^*$. From left to right, temperatures are $T=1.0, 0.6, 0.5, 0.4, 0.375, 0.35, 0.3, 0.29$.}
\label{FQT-525}
\end{figure}

The presence of subdiffusivity is a hint of a closeby higher order singularity, but in order to provide a more convincing proof of its existence, we now look at the behaviour of the density auto-correlations functions.
A distinctive feature is the presence of a pure logarithmic regime for a certain wave-vector $q^*$, where the second-order term of the asymptotic expansion in Eq.~\ref{eq:log} vanishes. Below and above $q^*$ the data should display a typical concave-to-convex transition. To visualise this behaviour one should be close enough to the $A_3^L$ point, but far enough from the liquid-glass transition in order to avoid that the final two-step decay covers most of the time-window and preempts the observation of the logarithmic behaviour. We identify at this $\phi$ the optimal temperature obeying these requisites as $T=0.375$, for which we show the collective normalised density auto-correlations functions $\Phi^{AA}_q(t)$ as a function of wave-vector in Fig.~\ref{FQT-525}(a). Indeed, at this $T$ we are able to identify $q^*\sigma_{AA}\approx 7$ where the decay of the correlators is purely logarithmic.  Across $q^*$, the concave-to-convex transition in the shape of $\Phi_q(t)$ with time is found. 
In Fig.~\ref{FQT-525}(b), the $T$-dependence of the correlators at fixed $q=q^*$ is shown. It is clear that with further decreasing $T$ the system approaches the liquid-glass transition, so that the signal of the logarithmic decay gets lost. Indeed, the range where the logarithmic dependence (dashed line) is valid shrinks upon reducing $T$. 
The evidence reported so far points to the existence of a higher order singularity in the vicinity of the explored path. While we cannot probe its exact location, since the system falls out of equilibrium before this can be accessed, it seems to be located within, but not too far inside, the glassy region, compatibly with MCT predictions of a disconnected glass-glass line.


Next we investigate the presence of the second singularity $A_3^H$. To approach it, we monitor the isotherm $T=0.5$ and check the dynamical behaviour with increasing $\phi$. We recall that the endpoint should be located at $\phi\sim0.6$, while the system should become glassy around $\phi \sim 0.593$, as estimated by the power law fits of $D_A$. We repeat the same analysis as done for $A_3^L$ and we find very similar results (not shown). Indeed, also in this case, we observe subdiffusive regime and logarithmic dynamics centred around a similar value of $q^*$.
Also in this case, the interference of the liquid-glass line does not allow us to probe the anomalous time window for a significant amount of time. To shed light on this point, simulations need be performed for a system with a larger $\Delta$ where the glass-glass line merges with the liquid-glass line so that the anomalous dynamics can be approached in equilibrium. This investigation is currently in progress.

\subsection{Non-ergodicity parameters from MCT and simulations}
\label{sec:fq}
In this section, we show the behaviour of the non-ergodicity parameters calculated within MCT and extracted from the fits of the density auto-correlation functions in the simulations.

\begin{figure}[ht]
\begin{center}
\includegraphics[scale=0.3, angle=0]{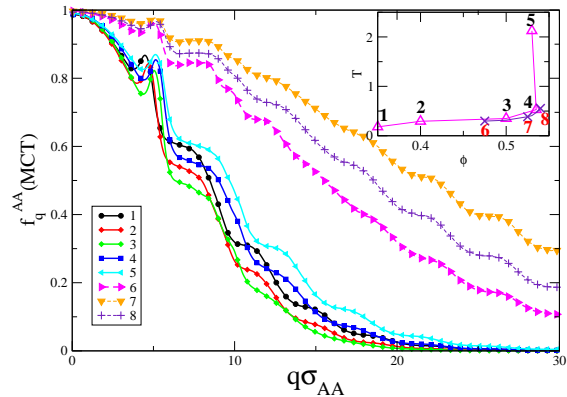}
\caption{Critical non-ergodicity parameters for the A species calculated within MCT along the liquid-glass (curves labeled from 1 to 5) and along the glass-glass (curves labeled from 6 to 8) lines. The corresponding state points and their position on the MCT lines are reported in the inset: a non-monotonic behaviour with increasing $\phi$ is observed for both sets of data.}
\label{fig:fq-MCT}
\end{center}
\end{figure}

We start by showing in Fig.~\ref{fig:fq-MCT} the `critical' non-ergodicity parameters $f_q^{AA}(MCT)$ for the A species along the liquid-glass and the glass-glass lines, calculated within MCT. The state points at which $f_q^{AA}(MCT)$ are evaluated are shown and numbered in the inset of Fig. ~\ref{fig:fq-MCT}.
We find that along the liquid-glass line the theoretical $f_q^{AA}(MCT)$ follows two different types of behaviour:
for $\phi\leq 0.50$, $f_q^{AA}(MCT)$ at first decreases, showing a shift of the peaks to larger wave vectors (from state point 1 to 3), while for $\phi>0.50$ at all temperatures ---below and above the reentrance--- $f_q^{AA}(MCT)$ maintains its peaks and only grows around them in a continuous way (from state point 3 to 5). We note that the $f_q^{AA}(MCT)$ of the lowest and highest $T$ (states 1 and 5) corresponds to two glasses made, respectively, by effective hard spheres of diameter $\sigma+\Delta$ and hard spheres of diameter $\sigma$. It follows that the two $f_q^{AA}(MCT)$ can be superimposed on top of each other by simply readjusting the diameters. In between these two limits, the system experiences the competition between the two length scales, which results in a non-trivial and non-monotonic behaviour of the non-ergodicity parameters.  The crossover between the two regimes arises in a region that cannot be associated with any of the two higher order singularities.

Furthermore, we also show in Fig.~\ref{fig:fq-MCT} the evolution of $f_q^{AA}(MCT)$'s for the `disconnected' glass, occurring upon crossing the glass-glass line. In this case the non-ergodicity parameters are much larger and longer-ranged in $q$, indicating more tightly-caged glasses, in analogy with previous observations of repulsive glasses for star polymer mixtures\cite{MayerMacro}. In the present case, the cage size is approximately one half of that of the first glass, as it can be estimated by the $q$-range of the non-ergodicity parameters. We observe a non-monotonic behaviour when moving along the glass-glass line from low to high density, similarly to what observed for the liquid-glass line.

We now extract the non-ergodicity parameters by fitting the density auto-correlation functions close to the arrest line via the stretched exponentials of Eq.~(\ref{eq:stretch}). For simplicity,  again we focus only on species $A$, since the results are formally identical for both species.

\begin{figure}[h]
\includegraphics[scale=0.3, angle=0]{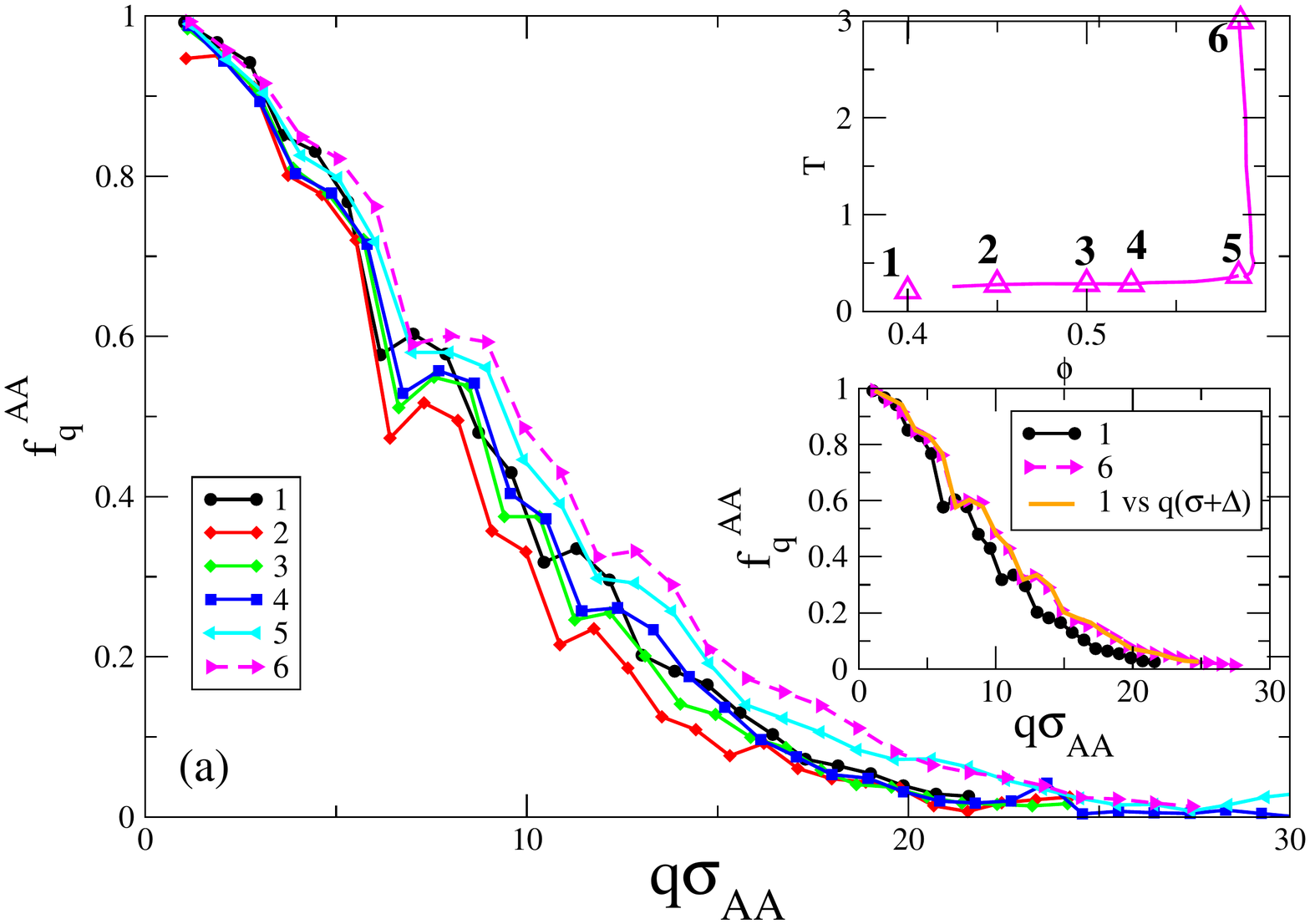}
 \includegraphics[scale=0.3, angle=0]{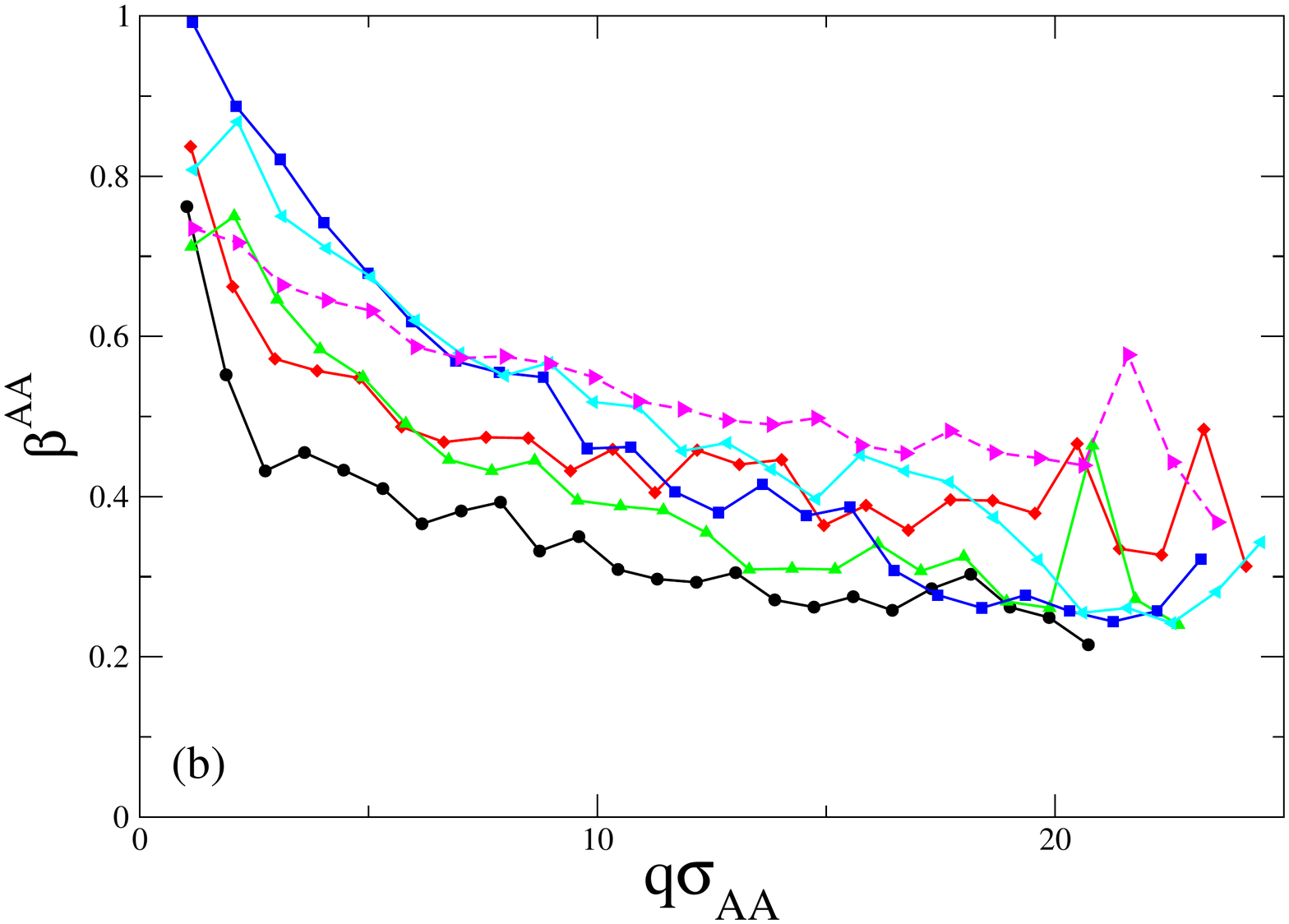}
\caption{(a): Non-ergodicity parameters obtained from simulations, fitting the density auto-correlation functions with stretched exponentials, for the state points reported in the upper inset. The behaviour along the liquid-glass line is strikingly similar to that of MCT predictions, reported in Fig.~\protect\ref{fig:fq-MCT}.  Lower inset:  $f_q^{AA}$ for low $\phi$ (and low $T$) (state point 1) is identical to that of high $\phi$ (state point 6) upon a rescaling by the effective diameter $\sigma+\Delta$; (b) Stretching exponents $\beta^{AA}$ obtained from the fits as a function of wave vector for the same state points considered in (a).}
\label{fig:fq-simu}
\end{figure}

We show in Fig.~\ref{fig:fq-simu} the behaviour of $f_q^{AA}$ along the liquid-glass line (for the state points labeled in the upper inset). We find a behaviour that is remarkably similar to the MCT predictions. Indeed, again we find that the non-ergodicity parameter shows a non-monotonic behaviour (at fixed $q$) with increasing $\phi$.
While at low $\phi$ and low $T$, $f_q^{AA}$ is compatible with that of a HS with effective diameter $\sigma+\Delta$, it decreases at first with a shift of the peaks at larger $q$ for intermediate densities. Then at a certain point, that we can roughly estimate as $\phi\approx 0.50$, the peaks do not move further in $q$ while $f_q^{AA}$ starts to increase in magnitude. This continues until it behaves as the $f_q$ typical for HS of diameter $\sigma$. This interpretation is reinforced by the fact that we can perfectly scale  the low-$\phi$ $f_q^{AA}$ on top of the high $\phi$-one by simply scaling it for the effective diameter (see lower inset). In addition, the behaviour of the stretching exponents obtained from the fits seems to indicate an increase of $\beta^{AA}$ as the system moves along the liquid-glass line with increasing $\phi$. Hence glasses found at lower $T$ are considerably more stretched ($\beta^{AA} \lesssim 0.4$) than those found at higher $\phi$. 

Finally we could not estimate numerically the non-ergodicity parameters close the disconnected glass-glass transition, due to the fact that this is buried inside the glass phase. Hence, we do not have a numerical analogue of curves (6-8) in Fig.~\ref{fig:fq-MCT}. 


\section{Discussion and Conclusions}
\label{sec:discuss}
In this work we have reported an extensive investigation by simulations and theory of the SS model. Despite the simplicity of the model, its dynamical behaviour close to the glass transition had largely remained unexplored, while its thermodynamic phase diagram showing the presence of many crystal phases has been addressed in a number of publications \cite{Gernot08,Prestipino09, Ziherl01}. The recent predictions by MCT of a unique behaviour of the glass transition for the SS model \cite{Sperl2010} is the motivation for the present work. Given the amount of effort requested to elucidate the various aspects of both liquid-glass transition and glass-glass lines, we have focused on a single value of the width of the SS model, referring to future studies to address the dependence on $\Delta$ of the results presented here.

Our investigation initially focused on searching for the so-called diffusion anomalies, i.e. a local maximum in the self-diffusion coefficient, both upon compressing the system and upon cooling it, as predicted by MCT. We performed the simulations down to a $T$-range at the limit of today computational capabilities and we did not detect the presence of an anomaly upon compression/expansion in contrast to theoretical results, as well as with other studies of core-softened potentials  \cite{Kumar05,core11,Malescio0a,Malescio0b}. However, none of these previous studies involved a potential with sharply-defined length scales such as the SS, so that this might be the reason why a different behaviour is observed. Also, it remains to be established whether such different models share the intriguing glassy properties of the one-component SS model, as detailed MCT studies have not been performed. To rule out in the SS model the presence of a diffusivity maximum in density we will need to explore other values of $\Delta$ in the future. Our preliminary investigations in this direction have also been negative in this respect, but we believe that we need to perform additional simulations for significantly slow state points in order to resolve this controversial aspect.

The role of the SS width is also crucial within MCT, because it controls the position of the glass-glass line and its endpoints with respect to the liquid-glass line. The present case was predicted  to show two glass-glass lines merging from the liquid-glass one by RY-MCT for the 
monodisperse SS \cite{Sperl2010}. However, we have repeated the MCT calculations for the binary mixture investigated by MD simulations, which avoided the onset of crystallisation, and  found a different topology. Namely, a disconnected glass-glass transition lying all inside the 
glassy region with two endpoints, i.e. a scenario that RY-MCT attributed to lower values of $\Delta$. Such a shift in the control parameters is expected when comparing with MCT, although for the SW model there was a remarkable good agreement between theory and simulations at the same $\Delta$ \cite{Sci03a}. Also no reentrance in $\phi$ was detected within the binary MCT calculations, in agreement with the simulation results. The disconnected glass-glass topology seems to be confirmed by simulations, which have shown the presence of a regime of logarithmic decay of the correlators and of subdiffusive increase of the MSD, in agreement with the predictions close to a higher order singularity.  Also the presence of the glass-glass line and its special endpoints, called $A_3^L$ and $A_3^H$, is compatible with the presented results. However, to really confirm the predictions, it will be desirable in the future to approach these points in equilibrium, as previously done for the SW model, by a fine tuning of $\Delta$. Again, we plan to undertake these investigations in the near future. 


Despite the absence of clear results about the glass-glass transition, we have found evidence that the non-ergodicity parameters along the liquid-glass line vary in a non-monotonic fashion, in very good agreement with the theory. We have shown that the glass found at low/intermediate $\phi$ (e.g. $\phi=0.40$) and low $T$ is identical to that found at high $\phi$ (and almost $T$-independent) when a scaling of the effective diameter is performed to take into account the effect of the shoulder width. Thus the system clearly displays two {\it identical} glasses, both of HS type, driven solely by excluded volume, and differing between themselves only by a change of length scale. However, the interplay between these two glasses is highly non-trivial, giving rise to anomalous behaviour in the dynamics, although not in the form of a local maximum in the diffusion coefficient, but in the form of clear subdiffusive regime in the MSD and logarithmic dynamics of the density correlators. This is enhanced at a length scale that is compatible with that always associated as the main responsible for the hard-sphere transition, i.e. the nearest-neighbour length.  It is a remarkable finding that this simple physics is capable of producing such non-trivial and unexpected dynamical behaviour. These results clearly show that, in order to find higher order MCT singularities, it is not needed to rely on two different physical ingredients, such as attraction and repulsion, but competing isotropic repulsions are sufficient.

Finally another aspect that is very peculiar to the present study is the fact that at low $\phi$ we can approach the glass transition down to very low temperatures, without intervening crystallisation or phase separation. This was previously achieved in other systems with directional attraction, where however the dynamics at low $T$ was found to be dominated by bonding processes, showing an Arrhenius dependence. Here on the contrary, bonding is not present and the dynamics seems to always retain the fragile character of HS systems. The absence of phase separation at low $T$ and $\phi$ for the present system thus offers the unexplored possibility to investigate the glass transition at  $T\rightarrow 0$ for $\phi\rightarrow \phi_g [\sigma/(\sigma+\Delta)]^3$ for which current work is underway.

\section{Acknowledgments}
We acknowledge helpful discussions with Matthias Sperl, Sergey V. Buldyrev, Pradeep Kumar and H. Eugene Stanley. This work was supported by EU through ITN-234810-COMPLOIDS  and ERC-226207-PATCHYCOLLOIDS, and by MIUR-FIRB  ANISOFT (RBFR125H0M).


\begin{thebibliography}{50}
\expandafter\ifx\csname natexlab\endcsname\relax\def\natexlab#1{#1}\fi
\expandafter\ifx\csname bibnamefont\endcsname\relax
  \def\bibnamefont#1{#1}\fi
\expandafter\ifx\csname bibfnamefont\endcsname\relax
  \def\bibfnamefont#1{#1}\fi
\expandafter\ifx\csname citenamefont\endcsname\relax
  \def\citenamefont#1{#1}\fi
\expandafter\ifx\csname url\endcsname\relax
  \def\url#1{\texttt{#1}}\fi
\expandafter\ifx\csname urlprefix\endcsname\relax\def\urlprefix{URL }\fi
\providecommand{\bibinfo}[2]{#2}
\providecommand{\eprint}[2][]{\url{#2}}

\bibitem[{\citenamefont{Pusey and {van Megen}}(1987)}]{Pus87a}
\bibinfo{author}{\bibfnamefont{P.~N.} \bibnamefont{Pusey}} \bibnamefont{and}
  \bibinfo{author}{\bibfnamefont{W.}~\bibnamefont{{van Megen}}},
  \bibinfo{journal}{Phys. Rev. Lett.} \textbf{\bibinfo{volume}{59}},
  \bibinfo{pages}{2083} (\bibinfo{year}{1987}).

\bibitem[{\citenamefont{{van Megen} and {Underwood}}(1993)}]{vanM93}
\bibinfo{author}{\bibfnamefont{W.}~\bibnamefont{{van Megen}}} \bibnamefont{and}
  \bibinfo{author}{\bibfnamefont{S.~M.} \bibnamefont{{Underwood}}},
  \bibinfo{journal}{Phys. Rev. Lett.} \textbf{\bibinfo{volume}{70}},
  \bibinfo{pages}{2766} (\bibinfo{year}{1993}).

\bibitem[{\citenamefont{G\"otze}(1991)}]{goetze}
\bibinfo{author}{\bibfnamefont{W.}~\bibnamefont{G\"otze}},
  \emph{\bibinfo{title}{Liquids, Freezing and the Glass Transition}}
  (\bibinfo{publisher}{North-Holland Amsterdam}, \bibinfo{year}{1991}), pp.
  \bibinfo{pages}{287--503}.

\bibitem[{\citenamefont{G\"{o}tze and Sperl}(2002)}]{Gotzesperl}
\bibinfo{author}{\bibfnamefont{W.}~\bibnamefont{G\"{o}tze}} \bibnamefont{and}
  \bibinfo{author}{\bibfnamefont{M.}~\bibnamefont{Sperl}},
  \bibinfo{journal}{Phys. Rev. E} \textbf{\bibinfo{volume}{66}},
  \bibinfo{pages}{011405} (\bibinfo{year}{2002}).

\bibitem[{\citenamefont{Pham et~al.}(2002)\citenamefont{Pham, Puertas,
  Bergenholtz, Egelhaaf, Moussa{\" i}d, Pusey, Schofield, Cates, Fuchs, and
  Poon}}]{Pha02a}
\bibinfo{author}{\bibfnamefont{K.~N.} \bibnamefont{Pham}},
  \bibinfo{author}{\bibfnamefont{A.~M.} \bibnamefont{Puertas}},
  \bibinfo{author}{\bibfnamefont{J.}~\bibnamefont{Bergenholtz}},
  \bibinfo{author}{\bibfnamefont{S.~U.} \bibnamefont{Egelhaaf}},
  \bibinfo{author}{\bibfnamefont{A.}~\bibnamefont{Moussa{\" i}d}},
  \bibinfo{author}{\bibfnamefont{P.~N.} \bibnamefont{Pusey}},
  \bibinfo{author}{\bibfnamefont{A.~B.} \bibnamefont{Schofield}},
  \bibinfo{author}{\bibfnamefont{M.~E.} \bibnamefont{Cates}},
  \bibinfo{author}{\bibfnamefont{M.}~\bibnamefont{Fuchs}}, \bibnamefont{and}
  \bibinfo{author}{\bibfnamefont{W.~C.~K.} \bibnamefont{Poon}},
  \bibinfo{journal}{Science} \textbf{\bibinfo{volume}{296}},
  \bibinfo{pages}{104} (\bibinfo{year}{2002}).

\bibitem[{\citenamefont{Fabbian et~al.}(1999)\citenamefont{Fabbian, G{\" o}tze,
  Sciortino, Tartaglia, and Thiery}}]{Fab99a}
\bibinfo{author}{\bibfnamefont{L.}~\bibnamefont{Fabbian}},
  \bibinfo{author}{\bibfnamefont{W.}~\bibnamefont{G{\" o}tze}},
  \bibinfo{author}{\bibfnamefont{F.}~\bibnamefont{Sciortino}},
  \bibinfo{author}{\bibfnamefont{P.}~\bibnamefont{Tartaglia}},
  \bibnamefont{and} \bibinfo{author}{\bibfnamefont{F.}~\bibnamefont{Thiery}},
  \bibinfo{journal}{Phys. Rev. E} \textbf{\bibinfo{volume}{59}},
  \bibinfo{pages}{1347} (\bibinfo{year}{1999}).

\bibitem[{\citenamefont{Bergenholtz and Fuchs}(1999)}]{Ber99a}
\bibinfo{author}{\bibfnamefont{J.}~\bibnamefont{Bergenholtz}} \bibnamefont{and}
  \bibinfo{author}{\bibfnamefont{M.}~\bibnamefont{Fuchs}},
  \bibinfo{journal}{Phys. Rev. E} \textbf{\bibinfo{volume}{59}},
  \bibinfo{pages}{5706} (\bibinfo{year}{1999}).

\bibitem[{\citenamefont{Dawson et~al.}(2001)\citenamefont{Dawson, Foffi, Fuchs,
  G\"otze, Sciortino, Sperl, Tartaglia, Voigtmann, and Zaccarelli}}]{Daw00a}
\bibinfo{author}{\bibfnamefont{K.~A.} \bibnamefont{Dawson}},
  \bibinfo{author}{\bibfnamefont{G.}~\bibnamefont{Foffi}},
  \bibinfo{author}{\bibfnamefont{M.}~\bibnamefont{Fuchs}},
  \bibinfo{author}{\bibfnamefont{W.}~\bibnamefont{G\"otze}},
  \bibinfo{author}{\bibfnamefont{F.}~\bibnamefont{Sciortino}},
  \bibinfo{author}{\bibfnamefont{M.}~\bibnamefont{Sperl}},
  \bibinfo{author}{\bibfnamefont{P.}~\bibnamefont{Tartaglia}},
  \bibinfo{author}{\bibfnamefont{T.}~\bibnamefont{Voigtmann}},
  \bibnamefont{and}
  \bibinfo{author}{\bibfnamefont{E.}~\bibnamefont{Zaccarelli}},
  \bibinfo{journal}{Phys. Rev. E} \textbf{\bibinfo{volume}{63}},
  \bibinfo{pages}{011401} (\bibinfo{year}{2001}).

\bibitem[{\citenamefont{Sciortino}(2002)}]{Sciortino2002b}
\bibinfo{author}{\bibfnamefont{F.}~\bibnamefont{Sciortino}},
  \bibinfo{journal}{Nature Materials} \textbf{\bibinfo{volume}{1}},
  \bibinfo{pages}{145} (\bibinfo{year}{2002}).

\bibitem[{\citenamefont{Puertas et~al.}(2002)\citenamefont{Puertas, Fuchs, and
  Cates}}]{Pue02a}
\bibinfo{author}{\bibfnamefont{A.~M.} \bibnamefont{Puertas}},
  \bibinfo{author}{\bibfnamefont{M.}~\bibnamefont{Fuchs}}, \bibnamefont{and}
  \bibinfo{author}{\bibfnamefont{M.~E.} \bibnamefont{Cates}},
  \bibinfo{journal}{Phys. Rev. Lett.} \textbf{\bibinfo{volume}{88}},
  \bibinfo{pages}{098301} (\bibinfo{year}{2002}).

\bibitem[{\citenamefont{Zaccarelli et~al.}(2002)\citenamefont{Zaccarelli,
  Foffi, Dawson, Buldyrev, Sciortino, and Tartaglia}}]{Zac02a}
\bibinfo{author}{\bibfnamefont{E.}~\bibnamefont{Zaccarelli}},
  \bibinfo{author}{\bibfnamefont{G.}~\bibnamefont{Foffi}},
  \bibinfo{author}{\bibfnamefont{K.~A.} \bibnamefont{Dawson}},
  \bibinfo{author}{\bibfnamefont{S.~V.} \bibnamefont{Buldyrev}},
  \bibinfo{author}{\bibfnamefont{F.}~\bibnamefont{Sciortino}},
  \bibnamefont{and}
  \bibinfo{author}{\bibfnamefont{P.}~\bibnamefont{Tartaglia}},
  \bibinfo{journal}{Phys. Rev. E} \textbf{\bibinfo{volume}{66}},
  \bibinfo{pages}{041402} (\bibinfo{year}{2002}).

\bibitem[{\citenamefont{Zaccarelli and Poon}(2009)}]{ZaccarelliPNAS}
\bibinfo{author}{\bibfnamefont{E.}~\bibnamefont{Zaccarelli}} \bibnamefont{and}
  \bibinfo{author}{\bibfnamefont{W.~C.~K.} \bibnamefont{Poon}},
  \bibinfo{journal}{Proc. Natl. Acad. Sci. U.S.A.}
  \textbf{\bibinfo{volume}{106}}, \bibinfo{pages}{15203}
  (\bibinfo{year}{2009}).

\bibitem[{\citenamefont{Eckert and Bartsch}(2002)}]{Eck02a}
\bibinfo{author}{\bibfnamefont{T.}~\bibnamefont{Eckert}} \bibnamefont{and}
  \bibinfo{author}{\bibfnamefont{E.}~\bibnamefont{Bartsch}},
  \bibinfo{journal}{Phys. Rev. Lett.} \textbf{\bibinfo{volume}{89}},
  \bibinfo{pages}{125701} (\bibinfo{year}{2002}).

\bibitem[{\citenamefont{Foffi et~al.}(2002)\citenamefont{Foffi, Dawson,
  Buldyrev, Sciortino, Zaccarelli, and Tartaglia}}]{Fof02a}
\bibinfo{author}{\bibfnamefont{G.}~\bibnamefont{Foffi}},
  \bibinfo{author}{\bibfnamefont{K.~A.} \bibnamefont{Dawson}},
  \bibinfo{author}{\bibfnamefont{S.~V.} \bibnamefont{Buldyrev}},
  \bibinfo{author}{\bibfnamefont{F.}~\bibnamefont{Sciortino}},
  \bibinfo{author}{\bibfnamefont{E.}~\bibnamefont{Zaccarelli}},
  \bibnamefont{and}
  \bibinfo{author}{\bibfnamefont{P.}~\bibnamefont{Tartaglia}},
  \bibinfo{journal}{Phys. Rev. E} \textbf{\bibinfo{volume}{65}},
  \bibinfo{pages}{050802} (\bibinfo{year}{2002}).

\bibitem[{\citenamefont{Likos et~al.}(1998)\citenamefont{Likos, L{\"o}wen,
  Watzlawek, Abbas, Juchnischke, Allgaier, and Richter}}]{Lik98c}
\bibinfo{author}{\bibfnamefont{C.~N.} \bibnamefont{Likos}},
  \bibinfo{author}{\bibfnamefont{H.}~\bibnamefont{L{\"o}wen}},
  \bibinfo{author}{\bibfnamefont{M.}~\bibnamefont{Watzlawek}},
  \bibinfo{author}{\bibfnamefont{B.}~\bibnamefont{Abbas}},
  \bibinfo{author}{\bibfnamefont{O.}~\bibnamefont{Juchnischke}},
  \bibinfo{author}{\bibfnamefont{J.}~\bibnamefont{Allgaier}}, \bibnamefont{and}
  \bibinfo{author}{\bibfnamefont{D.}~\bibnamefont{Richter}},
  \bibinfo{journal}{Phys. Rev. Lett.} \textbf{\bibinfo{volume}{80}},
  \bibinfo{pages}{4450} (\bibinfo{year}{1998}).

\bibitem[{\citenamefont{Foffi et~al.}(2003)\citenamefont{Foffi, Sciortino,
  Tartaglia, Zaccarelli, {Lo Verso}, Reatto, Dawson, and Likos}}]{Fof03a}
\bibinfo{author}{\bibfnamefont{G.}~\bibnamefont{Foffi}},
  \bibinfo{author}{\bibfnamefont{F.}~\bibnamefont{Sciortino}},
  \bibinfo{author}{\bibfnamefont{P.}~\bibnamefont{Tartaglia}},
  \bibinfo{author}{\bibfnamefont{E.}~\bibnamefont{Zaccarelli}},
  \bibinfo{author}{\bibfnamefont{F.}~\bibnamefont{{Lo Verso}}},
  \bibinfo{author}{\bibfnamefont{L.}~\bibnamefont{Reatto}},
  \bibinfo{author}{\bibfnamefont{K.~A.} \bibnamefont{Dawson}},
  \bibnamefont{and} \bibinfo{author}{\bibfnamefont{C.~N.} \bibnamefont{Likos}},
  \bibinfo{journal}{Phys. Rev. Lett.} \textbf{\bibinfo{volume}{90}},
  \bibinfo{pages}{238301} (\bibinfo{year}{2003}).

\bibitem[{\citenamefont{Mayer et~al.}(2008)\citenamefont{Mayer, Zaccarelli,
  Stiakakis, Likos, Sciortino, Munam, Gauthier, Hadjichristidis, Iatrou,
  Tartaglia et~al.}}]{MayerNature}
\bibinfo{author}{\bibfnamefont{C.}~\bibnamefont{Mayer}},
  \bibinfo{author}{\bibfnamefont{E.}~\bibnamefont{Zaccarelli}},
  \bibinfo{author}{\bibfnamefont{E.}~\bibnamefont{Stiakakis}},
  \bibinfo{author}{\bibfnamefont{C.~N.} \bibnamefont{Likos}},
  \bibinfo{author}{\bibfnamefont{F.}~\bibnamefont{Sciortino}},
  \bibinfo{author}{\bibfnamefont{A.}~\bibnamefont{Munam}},
  \bibinfo{author}{\bibfnamefont{M.}~\bibnamefont{Gauthier}},
  \bibinfo{author}{\bibfnamefont{N.}~\bibnamefont{Hadjichristidis}},
  \bibinfo{author}{\bibfnamefont{H.}~\bibnamefont{Iatrou}},
  \bibinfo{author}{\bibfnamefont{P.}~\bibnamefont{Tartaglia}},
  \bibnamefont{et~al.}, \bibinfo{journal}{Nat. Mater.}
  \textbf{\bibinfo{volume}{7}}, \bibinfo{pages}{780} (\bibinfo{year}{2008}).

\bibitem[{\citenamefont{Mayer et~al.}(2009)\citenamefont{Mayer, Sciortino,
  Likos, Tartaglia, Loewen, and Zaccarelli}}]{MayerMacro}
\bibinfo{author}{\bibfnamefont{C.}~\bibnamefont{Mayer}},
  \bibinfo{author}{\bibfnamefont{F.}~\bibnamefont{Sciortino}},
  \bibinfo{author}{\bibfnamefont{C.~N.} \bibnamefont{Likos}},
  \bibinfo{author}{\bibfnamefont{P.}~\bibnamefont{Tartaglia}},
  \bibinfo{author}{\bibfnamefont{H.}~\bibnamefont{Loewen}}, \bibnamefont{and}
  \bibinfo{author}{\bibfnamefont{E.}~\bibnamefont{Zaccarelli}},
  \bibinfo{journal}{Macromolecules} \textbf{\bibinfo{volume}{42}},
  \bibinfo{pages}{423} (\bibinfo{year}{2009}).

\bibitem[{\citenamefont{{Voigtmann}}(2011)}]{Voigt2011}
\bibinfo{author}{\bibfnamefont{T.}~\bibnamefont{{Voigtmann}}},
  \bibinfo{journal}{Europhys. Lett.} \textbf{\bibinfo{volume}{96}},
  \bibinfo{pages}{36006} (\bibinfo{year}{2011}).

\bibitem[{\citenamefont{Jagla}(1999)}]{Jagla}
\bibinfo{author}{\bibfnamefont{E.~A.} \bibnamefont{Jagla}},
  \bibinfo{journal}{J. Chem. Phys.} \textbf{\bibinfo{volume}{111}},
  \bibinfo{pages}{8980} (\bibinfo{year}{1999}).

\bibitem[{\citenamefont{Kumar et~al.}(2005)\citenamefont{Kumar, Buldyrev,
  Sciortino, Zaccarelli, and Stanley}}]{Kumar05}
\bibinfo{author}{\bibfnamefont{P.}~\bibnamefont{Kumar}},
  \bibinfo{author}{\bibfnamefont{S.~V.} \bibnamefont{Buldyrev}},
  \bibinfo{author}{\bibfnamefont{F.}~\bibnamefont{Sciortino}},
  \bibinfo{author}{\bibfnamefont{E.}~\bibnamefont{Zaccarelli}},
  \bibnamefont{and} \bibinfo{author}{\bibfnamefont{H.~E.}
  \bibnamefont{Stanley}}, \bibinfo{journal}{Phys. Rev. E.}
  \textbf{\bibinfo{volume}{72}}, \bibinfo{pages}{021501}
  (\bibinfo{year}{2005}).

\bibitem[{\citenamefont{Fomin et~al.}(2011)\citenamefont{Fomin, Tsiok, and
  Ryzhov}}]{core11}
\bibinfo{author}{\bibfnamefont{Y.~D.} \bibnamefont{Fomin}},
  \bibinfo{author}{\bibfnamefont{E.~N.} \bibnamefont{Tsiok}}, \bibnamefont{and}
  \bibinfo{author}{\bibfnamefont{V.~N.} \bibnamefont{Ryzhov}},
  \bibinfo{journal}{J. Chem. Phys.} \textbf{\bibinfo{volume}{135}},
  \bibinfo{pages}{124512} (\bibinfo{year}{2011}).

\bibitem[{\citenamefont{Xu et~al.}(2006)\citenamefont{Xu, Buldyrev, Angell, and
  Stanley}}]{Ramp}
\bibinfo{author}{\bibfnamefont{L.}~\bibnamefont{Xu}},
  \bibinfo{author}{\bibfnamefont{S.~V.} \bibnamefont{Buldyrev}},
  \bibinfo{author}{\bibfnamefont{C.~A.} \bibnamefont{Angell}},
  \bibnamefont{and} \bibinfo{author}{\bibfnamefont{H.~E.}
  \bibnamefont{Stanley}}, \bibinfo{journal}{Phys. Rev. E.}
  \textbf{\bibinfo{volume}{74}}, \bibinfo{pages}{031108}
  (\bibinfo{year}{2006}).

\bibitem[{\citenamefont{Netz et~al.}(2001)\citenamefont{Netz, Starr, Stanley,
  and Barbosa}}]{repulsive}
\bibinfo{author}{\bibfnamefont{P.~A.} \bibnamefont{Netz}},
  \bibinfo{author}{\bibfnamefont{F.~W.} \bibnamefont{Starr}},
  \bibinfo{author}{\bibfnamefont{H.~E.} \bibnamefont{Stanley}},
  \bibnamefont{and} \bibinfo{author}{\bibfnamefont{M.~C.}
  \bibnamefont{Barbosa}}, \bibinfo{journal}{J. Chem. Phys.}
  \textbf{\bibinfo{volume}{115}}, \bibinfo{pages}{344} (\bibinfo{year}{2001}).

\bibitem[{\citenamefont{Gallo and Sciortino}(2012)}]{Gallo2012}
\bibinfo{author}{\bibfnamefont{P.}~\bibnamefont{Gallo}} \bibnamefont{and}
  \bibinfo{author}{\bibfnamefont{F.}~\bibnamefont{Sciortino}},
  \bibinfo{journal}{Phys. Rev. Lett.} \textbf{\bibinfo{volume}{109}},
  \bibinfo{pages}{177801} (\bibinfo{year}{2012}).

\bibitem[{\citenamefont{Young and Alder}(1977)}]{cesium77}
\bibinfo{author}{\bibfnamefont{D.~A.} \bibnamefont{Young}} \bibnamefont{and}
  \bibinfo{author}{\bibfnamefont{B.~J.} \bibnamefont{Alder}},
  \bibinfo{journal}{Phys. Rev. Lett.} \textbf{\bibinfo{volume}{38}},
  \bibinfo{pages}{1213} (\bibinfo{year}{1977}).

\bibitem[{\citenamefont{Osterman et~al.}(2007)\citenamefont{Osterman, Babic,
  Poberaj, Dobnikar, and Ziherl}}]{Osterman07}
\bibinfo{author}{\bibfnamefont{N.}~\bibnamefont{Osterman}},
  \bibinfo{author}{\bibfnamefont{D.}~\bibnamefont{Babic}},
  \bibinfo{author}{\bibfnamefont{I.}~\bibnamefont{Poberaj}},
  \bibinfo{author}{\bibfnamefont{J.}~\bibnamefont{Dobnikar}}, \bibnamefont{and}
  \bibinfo{author}{\bibfnamefont{P.}~\bibnamefont{Ziherl}},
  \bibinfo{journal}{Phys. Rev. Lett.} \textbf{\bibinfo{volume}{99}},
  \bibinfo{pages}{248301} (\bibinfo{year}{2007}).

\bibitem[{\citenamefont{Duran}(1999)}]{duran}
\bibinfo{author}{\bibfnamefont{J.}~\bibnamefont{Duran}},
  \emph{\bibinfo{title}{Sands and Powders and Grains: An Introduction to the
  Physics of Granular Materials}} (\bibinfo{publisher}{Springer, New York},
  \bibinfo{year}{1999}).

\bibitem[{\citenamefont{Horbach}(2008)}]{horbach08}
\bibinfo{author}{\bibfnamefont{J.}~\bibnamefont{Horbach}}, \bibinfo{journal}{J.
  Phys. Condens. Matter} \textbf{\bibinfo{volume}{20}}, \bibinfo{pages}{244118}
  (\bibinfo{year}{2008}).

\bibitem[{\citenamefont{{Sperl} et~al.}(2010)\citenamefont{{Sperl},
  {Zaccarelli}, {Sciortino}, {Kumar}, and {Stanley}}}]{Sperl2010}
\bibinfo{author}{\bibfnamefont{M.}~\bibnamefont{{Sperl}}},
  \bibinfo{author}{\bibfnamefont{E.}~\bibnamefont{{Zaccarelli}}},
  \bibinfo{author}{\bibfnamefont{F.}~\bibnamefont{{Sciortino}}},
  \bibinfo{author}{\bibfnamefont{P.}~\bibnamefont{{Kumar}}}, \bibnamefont{and}
  \bibinfo{author}{\bibfnamefont{H.~E.} \bibnamefont{{Stanley}}},
  \bibinfo{journal}{Phys. Rev. Lett.} \textbf{\bibinfo{volume}{104}},
  \bibinfo{pages}{145701} (\bibinfo{year}{2010}).

\bibitem[{\citenamefont{Sciortino et~al.}(2003)\citenamefont{Sciortino,
  Tartaglia, and Zaccarelli}}]{Sci03a}
\bibinfo{author}{\bibfnamefont{F.}~\bibnamefont{Sciortino}},
  \bibinfo{author}{\bibfnamefont{P.}~\bibnamefont{Tartaglia}},
  \bibnamefont{and}
  \bibinfo{author}{\bibfnamefont{E.}~\bibnamefont{Zaccarelli}},
  \bibinfo{journal}{Phys. Rev. Lett.} \textbf{\bibinfo{volume}{91}},
  \bibinfo{pages}{268301} (\bibinfo{year}{2003}).

\bibitem[{\citenamefont{Zaccarelli et~al.}(2009)\citenamefont{Zaccarelli,
  Valeriani, Sanz, Poon, Cates, and Pusey}}]{Zac09a}
\bibinfo{author}{\bibfnamefont{E.}~\bibnamefont{Zaccarelli}},
  \bibinfo{author}{\bibfnamefont{C.}~\bibnamefont{Valeriani}},
  \bibinfo{author}{\bibfnamefont{E.}~\bibnamefont{Sanz}},
  \bibinfo{author}{\bibfnamefont{W.~C.~K.} \bibnamefont{Poon}},
  \bibinfo{author}{\bibfnamefont{M.~E.} \bibnamefont{Cates}}, \bibnamefont{and}
  \bibinfo{author}{\bibfnamefont{P.~N.} \bibnamefont{Pusey}},
  \bibinfo{journal}{Phys. Rev. Lett.} \textbf{\bibinfo{volume}{103}},
  \bibinfo{pages}{135704} (\bibinfo{year}{2009}).

\bibitem[{\citenamefont{Sperl}(2003)}]{sperl}
\bibinfo{author}{\bibfnamefont{M.}~\bibnamefont{Sperl}},
  \bibinfo{journal}{Phys. Rev. E} \textbf{\bibinfo{volume}{68}},
  \bibinfo{pages}{031405} (\bibinfo{year}{2003}).

\bibitem[{\citenamefont{Sperl}(2010)}]{Sperl0a}
\bibinfo{author}{\bibfnamefont{M.}~\bibnamefont{Sperl}},
  \bibinfo{journal}{Prog. Theor. Phys.} \textbf{\bibinfo{volume}{184}},
  \bibinfo{pages}{211} (\bibinfo{year}{2010}).

\bibitem[{Not({\natexlab{a}})}]{Note1}
\bibinfo{note}{The MCT equations can also be generalised to the case of
  Brownian Dynamics which is more realistic to describe colloidal suspensions.
  However, the long-time limit features and main predictions are not affected
  by the different microscopic dynamics.}

\bibitem[{\citenamefont{Hansen and McDonald}(2006)}]{hansen}
\bibinfo{author}{\bibfnamefont{J.~P.} \bibnamefont{Hansen}} \bibnamefont{and}
  \bibinfo{author}{\bibfnamefont{I.~R.} \bibnamefont{McDonald}},
  \emph{\bibinfo{title}{Theory of simple liquids}}
  (\bibinfo{publisher}{Academic Press, New York}, \bibinfo{year}{2006}),
  \bibinfo{edition}{3rd} ed.

\bibitem[{\citenamefont{Dawson et~al.}(2002)\citenamefont{Dawson, Foffi,
  McCullagh, Sciortino, Tartaglia, and Zaccarelli}}]{Daw02b}
\bibinfo{author}{\bibfnamefont{K.~A.} \bibnamefont{Dawson}},
  \bibinfo{author}{\bibfnamefont{G.}~\bibnamefont{Foffi}},
  \bibinfo{author}{\bibfnamefont{G.~D.} \bibnamefont{McCullagh}},
  \bibinfo{author}{\bibfnamefont{F.}~\bibnamefont{Sciortino}},
  \bibinfo{author}{\bibfnamefont{P.}~\bibnamefont{Tartaglia}},
  \bibnamefont{and}
  \bibinfo{author}{\bibfnamefont{E.}~\bibnamefont{Zaccarelli}},
  \bibinfo{journal}{J. Phys.: Condens. Matter} \textbf{\bibinfo{volume}{14}},
  \bibinfo{pages}{2223} (\bibinfo{year}{2002}).

\bibitem[{\citenamefont{{Sperl}}(2010)}]{Sperl2}
\bibinfo{author}{\bibfnamefont{M.}~\bibnamefont{{Sperl}}},
  \bibinfo{journal}{Prog. Theor. Phys. Supp.} \textbf{\bibinfo{volume}{184}},
  \bibinfo{pages}{209} (\bibinfo{year}{2010}).

\bibitem[{\citenamefont{Saija et~al.}(2009)\citenamefont{Saija, Prestipino, and
  Malescio}}]{Malescio0a}
\bibinfo{author}{\bibfnamefont{F.}~\bibnamefont{Saija}},
  \bibinfo{author}{\bibfnamefont{S.}~\bibnamefont{Prestipino}},
  \bibnamefont{and} \bibinfo{author}{\bibfnamefont{G.}~\bibnamefont{Malescio}},
  \bibinfo{journal}{Phys. Rev. E.} \textbf{\bibinfo{volume}{80}},
  \bibinfo{pages}{031502} (\bibinfo{year}{2009}).

\bibitem[{\citenamefont{Zaccarelli et~al.}(2006)\citenamefont{Zaccarelli,
  Saika-Voivod, Buldyrev, Moreno, Tartaglia, and Sciortino}}]{EZ2006}
\bibinfo{author}{\bibfnamefont{E.}~\bibnamefont{Zaccarelli}},
  \bibinfo{author}{\bibfnamefont{I.}~\bibnamefont{Saika-Voivod}},
  \bibinfo{author}{\bibfnamefont{S.~V.} \bibnamefont{Buldyrev}},
  \bibinfo{author}{\bibfnamefont{A.~J.} \bibnamefont{Moreno}},
  \bibinfo{author}{\bibfnamefont{P.}~\bibnamefont{Tartaglia}},
  \bibnamefont{and}
  \bibinfo{author}{\bibfnamefont{F.}~\bibnamefont{Sciortino}},
  \bibinfo{journal}{J. Chem. Phys.} \textbf{\bibinfo{volume}{124}},
  \bibinfo{pages}{124908} (\bibinfo{year}{2006}).

\bibitem[{\citenamefont{{Mayer} et~al.}(2010)\citenamefont{{Mayer},
  {Sciortino}, {Tartaglia}, and {Zaccarelli}}}]{sphericaljcp2}
\bibinfo{author}{\bibfnamefont{C.}~\bibnamefont{{Mayer}}},
  \bibinfo{author}{\bibfnamefont{F.}~\bibnamefont{{Sciortino}}},
  \bibinfo{author}{\bibfnamefont{P.}~\bibnamefont{{Tartaglia}}},
  \bibnamefont{and}
  \bibinfo{author}{\bibfnamefont{E.}~\bibnamefont{{Zaccarelli}}},
  \bibinfo{journal}{J. Phys. Condens. Matter} \textbf{\bibinfo{volume}{22}},
  \bibinfo{pages}{104110} (\bibinfo{year}{2010}).

\bibitem[{\citenamefont{B{\"o}hmer et~al.}(1993)\citenamefont{B{\"o}hmer, Ngai,
  Angell, and Plazek}}]{Boh93a}
\bibinfo{author}{\bibfnamefont{R.}~\bibnamefont{B{\"o}hmer}},
  \bibinfo{author}{\bibfnamefont{K.~L.} \bibnamefont{Ngai}},
  \bibinfo{author}{\bibfnamefont{C.~A.} \bibnamefont{Angell}},
  \bibnamefont{and} \bibinfo{author}{\bibfnamefont{D.~J.}
  \bibnamefont{Plazek}}, \bibinfo{journal}{J. Chem. Phys.}
  \textbf{\bibinfo{volume}{99}}, \bibinfo{pages}{4201} (\bibinfo{year}{1993}).

\bibitem[{Not({\natexlab{b}})}]{Note2}
\bibinfo{note}{Courtesy of M. Sperl}.

\bibitem[{\citenamefont{Lang et~al.}(1999)\citenamefont{Lang, Kahl, Likos,
  L\"{o}wen, and Watzlawek}}]{Lang99}
\bibinfo{author}{\bibfnamefont{A.}~\bibnamefont{Lang}},
  \bibinfo{author}{\bibfnamefont{G.}~\bibnamefont{Kahl}},
  \bibinfo{author}{\bibfnamefont{C.~N.} \bibnamefont{Likos}},
  \bibinfo{author}{\bibfnamefont{H.}~\bibnamefont{L\"{o}wen}},
  \bibnamefont{and}
  \bibinfo{author}{\bibfnamefont{M.}~\bibnamefont{Watzlawek}},
  \bibinfo{journal}{J. Phys. Condens. Matter} \textbf{\bibinfo{volume}{11}},
  \bibinfo{pages}{10143} (\bibinfo{year}{1999}).

\bibitem[{\citenamefont{Mayer et~al.}(2007)\citenamefont{Mayer, Stiakakis,
  Zaccarelli, Likos, Sciortino, Tartaglia, L\"owen, and Vlassopoulos}}]{Rheo}
\bibinfo{author}{\bibfnamefont{C.}~\bibnamefont{Mayer}},
  \bibinfo{author}{\bibfnamefont{E.}~\bibnamefont{Stiakakis}},
  \bibinfo{author}{\bibfnamefont{E.}~\bibnamefont{Zaccarelli}},
  \bibinfo{author}{\bibfnamefont{C.~N.} \bibnamefont{Likos}},
  \bibinfo{author}{\bibfnamefont{F.}~\bibnamefont{Sciortino}},
  \bibinfo{author}{\bibfnamefont{P.}~\bibnamefont{Tartaglia}},
  \bibinfo{author}{\bibfnamefont{H.}~\bibnamefont{L\"owen}}, \bibnamefont{and}
  \bibinfo{author}{\bibfnamefont{D.}~\bibnamefont{Vlassopoulos}},
  \bibinfo{journal}{Rheol. Acta} \textbf{\bibinfo{volume}{46}},
  \bibinfo{pages}{611} (\bibinfo{year}{2007}).

\bibitem[{\citenamefont{G{\"otze} and Voigtmann}(2003)}]{gotze03}
\bibinfo{author}{\bibfnamefont{W.}~\bibnamefont{G{\"otze}}} \bibnamefont{and}
  \bibinfo{author}{\bibfnamefont{T.}~\bibnamefont{Voigtmann}},
  \bibinfo{journal}{Phys. Rev. E} \textbf{\bibinfo{volume}{67}},
  \bibinfo{pages}{021502} (\bibinfo{year}{2003}).

\bibitem[{\citenamefont{Pauschenwein and Kahl}(2008)}]{Gernot08}
\bibinfo{author}{\bibfnamefont{G.~J.} \bibnamefont{Pauschenwein}}
  \bibnamefont{and} \bibinfo{author}{\bibfnamefont{G.}~\bibnamefont{Kahl}},
  \bibinfo{journal}{J. Chem. Phys.} \textbf{\bibinfo{volume}{129}},
  \bibinfo{pages}{174107} (\bibinfo{year}{2008}).

\bibitem[{\citenamefont{Prestipino et~al.}(2009)\citenamefont{Prestipino,
  Saija, and Malescio}}]{Prestipino09}
\bibinfo{author}{\bibfnamefont{S.}~\bibnamefont{Prestipino}},
  \bibinfo{author}{\bibfnamefont{F.}~\bibnamefont{Saija}}, \bibnamefont{and}
  \bibinfo{author}{\bibfnamefont{G.}~\bibnamefont{Malescio}},
  \bibinfo{journal}{Soft Matter} \textbf{\bibinfo{volume}{5}},
  \bibinfo{pages}{2795} (\bibinfo{year}{2009}).

\bibitem[{\citenamefont{Ziherl and Kamien}(2001)}]{Ziherl01}
\bibinfo{author}{\bibfnamefont{P.}~\bibnamefont{Ziherl}} \bibnamefont{and}
  \bibinfo{author}{\bibfnamefont{R.~D.} \bibnamefont{Kamien}},
  \bibinfo{journal}{J. Chem. Phys. B} \textbf{\bibinfo{volume}{105}},
  \bibinfo{pages}{42} (\bibinfo{year}{2001}).

\bibitem[{\citenamefont{Prestipino et~al.}(2010)\citenamefont{Prestipino,
  Saija, and Malescio}}]{Malescio0b}
\bibinfo{author}{\bibfnamefont{S.}~\bibnamefont{Prestipino}},
  \bibinfo{author}{\bibfnamefont{F.}~\bibnamefont{Saija}}, \bibnamefont{and}
  \bibinfo{author}{\bibfnamefont{G.}~\bibnamefont{Malescio}},
  \bibinfo{journal}{J. Chem. Phys.} \textbf{\bibinfo{volume}{133}},
  \bibinfo{pages}{144504} (\bibinfo{year}{2010}).

\end{thebibliography}

\end{document}